\documentclass[journal=jctcce,manuscript=article]{achemso}
\SectionNumbersOn

\usepackage{verbatim}
\usepackage[normalem]{ulem}
\usepackage{pdfpages}
\usepackage{xr}  %
\usepackage{xspace}
\usepackage{siunitx}
\DeclareSIUnit\angstrom{\text{\AA}}
\usepackage{subcaption}
\usepackage{graphicx}
\usepackage{xcolor}
\usepackage[version=4]{mhchem}
\usepackage{cleveref}

\newcommand{\pk}{\ensuremath{\mathrm{p}K}\xspace}
\newcommand{\pka}{\ensuremath{\mathrm{p}K_{\mathrm{a}}}\xspace}
\newcommand{\pkb}{\ensuremath{\mathrm{p}K_{\mathrm{b}}}\xspace}
\newcommand{\pkw}{\ensuremath{\mathrm{p}K_{\mathrm{w}}}\xspace}

\newcommand{\eg}{e.g.\xspace}
\newcommand{\ie}{i.e.\xspace}
\newcommand{\kT}{\ensuremath{k_\mathrm{B}T}\xspace}

\newcommand{\abr}[1]{\uppercase{#1}}
\newcommand{\cv}{\abr{cv}\xspace}
\newcommand{\dft}{\abr{dft}\xspace}
\newcommand{\nqe}{\abr{nqe}\xspace}
\newcommand{\mlip}{\abr{mlip}\xspace}
\newcommand{\pimd}{\abr{pimd}\xspace}
\newcommand{\pet}{\abr{pet}\xspace}
\newcommand{\ti}{\abr{ti}\xspace}
\newcommand{\md}{\abr{md}\xspace}
\newcommand{\fes}{\abr{fes}\xspace}
\newcommand{\pmf}{\abr{pmf}\xspace}
\newcommand{\mae}{\abr{mae}\xspace}
\newcommand{\gga}{\abr{gga}\xspace}
\newcommand{\cip}{\abr{cip}\xspace}
\newcommand{\remd}{\abr{remd}\xspace}

\newcommand{\gnn}{\abr{gnn}\xspace}

\crefname{equation}{Equation}{Equations}
\Crefname{equation}{Equation}{Equations}
\crefformat{equation}{Equation~(#2#1#3)}
\Crefformat{equation}{Equation~(#2#1#3)}
\crefname{figure}{Figure}{Figures}
\Crefname{figure}{Figure}{Figures}
\crefname{table}{Table}{Tables}
\Crefname{table}{Table}{Tables}
\crefname{section}{Section}{Sections}
\Crefname{section}{Section}{Sections}

\makeatletter
\AtBeginDocument{\let\LS@rot\@undefined}
\makeatother

\author{Qianjun Xu}
\affiliation{Laboratory of Computational Science and Modeling, Institut des Mat\'eriaux, \'Ecole Polytechnique F\'ed\'erale de Lausanne, 1015 Lausanne, Switzerland}
\author{Simin Liu}
\affiliation{Laboratory of Computational Science and Modeling, Institut des Mat\'eriaux, \'Ecole Polytechnique F\'ed\'erale de Lausanne, 1015 Lausanne, Switzerland}
\author{Cesare Malosso}
\affiliation{Laboratory of Computational Science and Modeling, Institut des Mat\'eriaux, \'Ecole Polytechnique F\'ed\'erale de Lausanne, 1015 Lausanne, Switzerland}
\author{Philip Loche}
\affiliation{Laboratory of Computational Science and Modeling, Institut des Mat\'eriaux, \'Ecole Polytechnique F\'ed\'erale de Lausanne, 1015 Lausanne, Switzerland}
\author{Michele Ceriotti}
\email{michele.ceriotti@epfl.ch}
\affiliation{Laboratory of Computational Science and Modeling, Institut des Mat\'eriaux, \'Ecole Polytechnique F\'ed\'erale de Lausanne, 1015 Lausanne, Switzerland}

\title{Quantitative modeling of acid-base equilibria}

\begin{document}

\begin{abstract}
Simulating acid-base equilibria from first principles is a tremendous challenge, as it requires modeling bond-breaking events in a complex hydrogen-bonding environment. The quantum mechanical behavior of the protons is often important, complicating simulations even further.
We show that by combining state-of-the-art machine-learning potentials, accelerated sampling of the dissociation process, and an efficient estimation of nuclear quantum corrections it is possible to reliably compute the dissociation constants of different species in water. 
We demonstrate the transferability of this workflow, applying it successfully to the autoionization of water and to the entire ladder of the aqueous dissociation of phosphoric acid. 
The definition of the dissociation constant, the choice of order parameters, sampling artifacts and finite-size effects, as well as quantum nuclear effects all play a non-trivial role in determining the outcome of the calculation, so that cutting corners is likely to introduce biases that can only be compensated by fortuitous cancellations of errors. 

\end{abstract}

\section{Introduction}
\label{sec:intro}

Acidity and basicity are among the most fundamental properties of chemical species in solution, and they influence a wide range of chemical and biological processes~\cite{retention_factor,buffer}.
They are particularly important for proteins~\cite{isoelectric,pli} and drug design~\cite{drug_pka}, given that approximately 30\% of the residues found in proteins are ionizable~\cite{wilsonAccuratelyPredictingProtein2023}, and that an estimated 95\% of drug compounds behave as weak acids or bases~\cite{wells1988pharmaceutical}.
Acidity and basicity are usually quantified in terms of the acid and base dissociation constants, \pka and \pkb, which for a conjugate acid-base pair in water are linked to the autoionization constant of the solvent, \pkw, by
\begin{equation}
    \pkw = \pka + \pkb.
    \label{eq:conjugate-pair}
\end{equation}
Data-driven models that predict \pka values from the molecular structure have reached a high level of accuracy, but their generalizability beyond the training domain and their interpretability remain limited~\cite{johnston2023epik,abarbanel2024qupkake,luo2024bridging,genzling2026development}.
Here we are concerned with the alternative bottom-up route of computing dissociation constants directly from atomistic simulations, which does not rely on the availability of experimental data for similar compounds, and which provides at the same time a microscopic picture of the dissociation and solvation mechanisms.

Modeling acid-base equilibria by molecular dynamics (\md) is far from trivial.
The equilibrium involves the breaking and formation of covalent bonds, so the potential that drives the dynamics must be reactive, and \textit{ab initio} methods such as density functional theory (\dft) remain prohibitively expensive on the time scales needed for a quantitative estimate of a free energy that can be as large as \SI{100}{kJ/mol}.
Dissociation is a rare event, and requires enhanced sampling techniques such as metadynamics~\cite{metad,wtm}, which in turn require the definition of collective variables (\cv{s}) that can follow the proton as it hops between water molecules by a Grotthuss-like mechanism~\cite{agmon1995grotthuss}.
The \cv{s} that have been introduced for this purpose~\cite{grifoni2019microscopic} typically require the convergence of a two-dimensional free energy surface (\fes), which adds substantially to the cost.
Finally, nuclear quantum effects (\nqe{s}) cannot be neglected: the quantum mechanical behavior of the proton is known to be important for a wide range of hydrogen-bonded systems~\cite{ceri+16cr,wang+14pnas,marsalekQuantumDynamicsSpectroscopy2017,ceri+13pnas,chengInitioThermodynamicsLiquid2019,stolteNuclearQuantumEffects2024}, and it affects both the dissociation of acids and bases~\cite{naoh_nqe,cl_nqe,gt_mispair} and the autoionization of water~\cite{thomsen2021nuclear,dasgupta2025nuclear}.
Replacing classical \md with path integral molecular dynamics (\pimd) multiplies the cost of every step by the number of replicas, which makes the combination of \textit{ab initio} \pimd and enhanced sampling impractical for anything but the simplest systems.

Perhaps because of these difficulties, the literature on the direct computation of dissociation constants is remarkably inconsistent in how the equilibrium constant is extracted from the simulation.
Some studies take the difference between the minima of the free energy profile in the associated and dissociated states~\cite{tummanapelli2015estimating,perez2018advances,daub2019ab,ramanModelingSolvationAcidity2022,joutsuka2022molecular}, others integrate the Boltzmann weight over the two regions of the \cv{}~\cite{thomsen2021nuclear,dasgupta2025nuclear}, and others yet rely on the reversible work theorem to relate the dissociation constant to the potential of mean force (\pmf) between the products~\cite{davies2002estimating,wang2019first,li2021accurate,calegari2023probing,li2026predictive}.
As we shall see, these estimators can differ by several \pk units for the same underlying free energy profile.
Nuclear quantum effects are treated just as inconsistently: some works compare classical simulations with experimental values extrapolated to a fictitious classical limit~\cite{wang2019first} or to deuterated water~\cite{li2026predictive}, while a few perform explicit \pimd simulations~\cite{thomsen2021nuclear,dasgupta2025nuclear}.
Since the magnitude of the quantum correction depends on the system and on the underlying potential, it is difficult to compare values across studies, and it is even more difficult to tell whether the agreement with experiment that is often reported for purely classical calculations~\cite{calegari2023probing,liu2023mechanistic} reflects the accuracy of the model or a fortuitous cancellation of errors.

Machine-learning interatomic potentials (\mlip{s}) remove the main obstacle to a clean protocol, as they reproduce the reference electronic-structure method at a small fraction of the cost~\cite{chengInitioThermodynamicsLiquid2019,Jacobs_2025}.
The recent availability of ``foundation'' \mlip{s} trained on very diverse datasets~\cite{yuan2026foundation,pet-mad,PET-MAD-1.5-2026,mace-polar,dpa4} reduces further the amount of system-specific \textit{ab initio} data that is needed, because a general-purpose model can be fine-tuned to a higher level of theory with a few thousand structures~\cite{pet-mad,radova2025fine,mattersim,tompa2026finetuning}.
The remaining cost of \nqe{s} can be reduced by computing the quantum correction to the free energy difference by thermodynamic integration (\ti) over the nuclear mass~\cite{habe-mano11jcp,ceri-mark13jcp,ross+16prl, chengInitioThermodynamicsLiquid2019}, which only requires converging the quantum kinetic energy in the two end states rather than a quantum \fes along the whole dissociation path.

In this work we combine these ingredients into a workflow for the calculation of absolute acid-base equilibria, and we discuss each step in the spirit of a practical guide, highlighting the choices that affect the outcome quantitatively.
First, we discuss the definition of the dissociation constant in the dilute limit and show how the reversible work theorem leads to an estimator that is insensitive to the details of the dissociation distance, once the potential of mean force has been calibrated against the asymptotic Coulomb interaction of the products, which also provides a natural finite-size correction.
Second, we introduce a single, signed dissociation distance that describes both the acidic and the basic dissociation of a solute, including the autoionization of the solvent, so that only a one-dimensional free-energy profile needs to be converged.
Third, we show how to use metadynamics to build a bias and replica exchange with a fixed bias to obtain converged Boltzmann weights, and how to correct the classical results including \nqe{s} by mass thermodynamic integration.
We use the autoionization of water as an illustrative example, and then demonstrate the transferability of the protocol on the entire dissociation ladder of phosphoric acid, computing six \pka and \pkb values of \ce{H3PO4} and of the amphiprotic and basic phosphate ions.
The outline of the paper is as follows.
\Cref{sec:theory} presents the theory, discussing for each step the shortcomings of the alternatives that have been used in the literature.
\Cref{sec:mlip} discusses the choice of the reference level of theory and of the \mlip, and presents the results for water and for phosphoric acid.
\Cref{sec:conclusions} draws our conclusions.

\section{Theory}
\label{sec:theory}

\subsection{What we compute}
\label{sec:what}

Let us consider a weak acid \ce{AH} in water, which dissociates according to
\begin{equation}
    \ce{AH + H2O <=> A^- + H3O^+},
    \label{eq:dissociation}
\end{equation}
and whose acid dissociation constant is defined, in the dilute limit and in terms of concentrations, as
\begin{equation}
    K_\mathrm{a} = \frac{[\ce{A^-}][\ce{H3O^+}]}{[\ce{AH}]\, c^\circ},
    \label{eq:ka-def}
\end{equation}
where $c^\circ=\SI{1}{mol/L}$ is the standard concentration, and $\pka = -\log_{10} K_\mathrm{a}$.
The basic dissociation \ce{B + H2O <=> BH^+ + OH^-} is defined analogously, and the autoionization of water \ce{2H2O <=> H3O^+ + OH^-} corresponds to $K_\mathrm{w}=[\ce{H3O^+}][\ce{OH^-}]/(c^\circ)^2$.
In all cases, the dissociation produces two species of opposite charge, and the question is how to relate the equilibrium constant to a quantity that can be sampled in a simulation with a single solute in a box of solvent.

The cleanest connection is provided by the reversible work theorem~\cite{chandler1987introduction}.
Let $r$ be the distance between the two products, and $g(r)$ their pair distribution function, normalized so that $g(r)\to 1$ at large separations, where the two ions do not interact.
The potential of mean force $W(r)=-\kT \ln g(r)$ is the reversible work needed to bring the two ions from infinite separation to a distance $r$.
In the dilute limit, the concentration of associated pairs is obtained by counting the configurations in which the two ions are closer than a cutoff distance $r_\mathrm{c}$, so that the association constant reads $\int_0^{r_\mathrm{c}} 4\pi r^2 g(r)\,\mathrm{d}r$, and the dissociation constant is its inverse,
\begin{equation}
    K = \frac{1}{c^\circ}\left\{\int_0^{r_\mathrm{c}} 4\pi r^2 \exp\left[-\beta W(r) \right] \mathrm{d}r\right\}^{-1},
    \label{eq:equilib-const}
\end{equation}
where $\beta=1/\kT$, and $c^\circ$ is converted into \SI{1660.5391}{\angstrom^{-3}}.
It is worth noting that for autoionization, the prefactor of \cref{eq:equilib-const} is $1/({c^\circ}^2V)$ instead of $1/c^\circ$, where $V$ is the volume of the simulation box, see \cref{SI-eq:pkw-final-result}.
Two features of \cref{eq:equilib-const} should be stressed.
First, the integral is dominated by the region where $W(r)$ is close to its minimum, \ie by the associated state.
As long as $r_\mathrm{c}$ is chosen beyond the barrier that separates bound and dissociated configurations, its precise value has a negligible effect on $K$, because the integrand is exponentially small there.
Second, and more importantly, the integral is a probability: it counts how likely the pair is to be found in the associated basin relative to the ideal, non-interacting state.
As such it does not depend on the details of how the separation $r$ is computed at short range.
Any dissociation distance that reduces to the ion separation at large distances and that identifies the same associated basin yields the same $K$, because a change of variable in the integral only redistributes the same probability mass along the axis.
This is not the case for estimators that read off the free energy at specific points, as we discuss below.

The delicate part of this approach lies in the normalization of $g(r)$.
A simulation delivers the free energy profile along the dissociation distance only up to an additive constant, and in the presence of the long-ranged Coulomb attraction between the ions $W(r)$ does not reach a plateau at any distance that is accessible in a periodic box of a few hundred molecules.
In practice, one has to determine the constant $C$ such that the simulated profile $W_\mathrm{sim}(r)$ satisfies
\begin{equation}
    W_\mathrm{sim}(r) - C \to 0 \qquad \text{for } r\to\infty,
    \label{eq:normalization}
\end{equation}
and use $W(r)=W_\mathrm{sim}(r)-C$ in \cref{eq:equilib-const}.
Since an error of $\kT\ln 10\approx\SI{5.7}{kJ/mol}$ in $C$ translates into an error of one full \pk unit, this step deserves as much attention as the sampling of the free energy profile itself.

Our approach to fix $C$ is to match the simulated \pmf to that of a pair of point charges at intermediate distances, well beyond the contact ion pair.
For two ions of charges $q_1$ and $q_2$ in a dielectric medium, the asymptotic \pmf is the screened Coulomb interaction, which is known analytically in the thermodynamic limit.
In a periodic box, however, the two ions also interact with their own periodic images and with those of their partner, and this interaction is not negligible for the box sizes that can be afforded in a first-principles-quality simulation.
The correct reference profile is the \pmf of the point charges in a periodic box of the same size and shape as the simulation cell, $q_L(r)$, computed with Ewald-type summation, for which we use the \texttt{torchpme} package~\cite{torchpme}.

There is one more subtlety.
The directions that are accessible at a distance $r$ are only those for which every Cartesian component is smaller than $L/2$: beyond $r=L/2$ the spherical shell is clipped by the cube, and its accessible area $A(r)$ is smaller than $4\pi r^2$.
A potential of mean force obtained from the simulated profile with the full spherical Jacobian therefore contains a spurious term $-k_\mathrm{B}T\ln[A(r)/4\pi r^2]$, that rises steeply beyond $L/2$.
The reference that the simulation should be matched to is thus $W_L(r) = q_L(r) - k_\mathrm{B}T\ln[A(r)/4\pi r^2]$.
The accessible fraction of the shell has a closed form: it is one up to $L/2$, then $3L/2r-2$ up to $L/\sqrt{2}$ (six spherical caps are cut off by the faces of the cube), and up to $L\sqrt{3}/2$, where the sphere no longer fits, it includes an inclusion-exclusion term for the overlap of pairs of caps along the edges.
The offset is then obtained as the average difference
\begin{equation}
    C = \left\langle W_\mathrm{sim}(r) - W_L(r) \right\rangle_{r > r_\mathrm{c}}
    \label{eq:offset}
\end{equation}
over the range of separations where the two profiles are expected to be parallel, and the finite-size effects on the equilibrium constant relevant to the periodic electrostatic and box geometry are absorbed into $C$ at no extra cost, because $W_L(r)$ approaches the infinite-system Coulomb interaction that defines the normalization in \cref{eq:normalization}.

In real simulations, bounds are usually set to confine systems within small $r$ region to enhance the sampling performance.
To avoid the artifacts caused by the bounds, the offset is calculated from $r_\mathrm{c}$ until \SI{4}{\angstrom} below where the bounds are set.
When the two products carry a net total charge, as is the case for the dissociation of an ion, the Ewald sum requires a neutralizing background charge that introduces a shift; we remove it by subtracting $E_\mathrm{W}(q_1+q_2)^2$ from $W_L(r)$, where $E_\mathrm{W}$ is the Wigner self-energy of a single point charge interacting with its periodic images and the neutralizing background, $q_1$ and $q_2$ are the charge carried by the dissociation products, respectively~\cite{hummer1997ion}.
It is worth stressing that this calibration works also when the \mlip that drives the simulation does not include explicit long-range electrostatics.
The matching in \cref{eq:offset} only needs a window of separations that is large enough for the two ions to be beyond the contact and solvent-separated pairs, but still within the receptive field of the model, where the \mlip has learned the effective, solvent-mediated interaction between the ions.
What is required is that the shape of the simulated \pmf follows the screened Coulomb interaction in that window, which is a useful sanity check on the potential in itself.

Two alternatives are commonly used in the literature, and both have shortcomings that are easy to overlook.
The first is to take the free energy difference between two points on the profile, typically the minima of the associated and dissociated states, and convert it into a \pk by dividing by $\kT\ln 10$~\cite{tummanapelli2015estimating,perez2018advances,daub2019ab,ramanModelingSolvationAcidity2022,joutsuka2022molecular}.
The value of a free energy profile at a point depends on the definition of the \cv{} through its Jacobian, and on the volume of configuration space that is mapped onto each value of the \cv{}, so that the result is sensitive to the choice of the dissociation distance and does not correspond to any well-defined standard state.
For the autoionization of water, we show in \cref{sec:water} that this estimator differs by 4.2 \pk units from that of \cref{eq:equilib-const} evaluated on the same free energy profile.
The second alternative is to integrate the Boltzmann weight of the free energy profile separately over the ``associated'' and the ``dissociated'' regions of the \cv{}~\cite{thomsen2021nuclear,dasgupta2025nuclear}, which is closer in spirit to \cref{eq:equilib-const} but makes the result depend strongly on the interval that is taken to represent the dissociated state.
Since the ions keep interacting at all separations that fit in the box, and the dissociated region does not correspond to a uniform density of free ions, the result depends on the box size in a way that cannot be corrected easily.
In a nutshell, the reversible work theorem moves all the arbitrariness into the single constant $C$, which can be determined from the known asymptotic form of the ion-ion interaction, including the effect of periodic boundary conditions.

\subsection{Defining dissociation distances}
\label{sec:cv}

The main challenge in defining the ion separation $r$ that enters \cref{eq:equilib-const} is that the products of the dissociation are not, in general, associated with a fixed set of atoms.
While the conjugate base \ce{A^-} of a molecular acid keeps its identity, the excess proton is a fluxional species that diffuses through the solvent by a Grotthuss mechanism~\cite{agmon1995grotthuss}, and so does the proton hole that constitutes \ce{OH^-}.
The first step is therefore to define the charge state of every potential proton-bearing site in the system.
We follow a smooth variation on the theme of the Voronoi-tessellation approach that assigns each hydrogen atom to the nearest oxygen~\cite{schw+01prl,grifoni2019microscopic}.
Each atom $j$ (in the following equation, all the hydrogen atoms) is partitioned among the ionizable sites $i$ (the oxygen atoms of water and of the acid) within a cutoff distance, with weights:
\begin{equation}
    n_{j\to i} = \frac{ e^{-\left| \mathbf{r}_j - \mathbf{r}_i \right|^2/2\sigma_j^2} }{ \sum_{k\in\mathcal{N}_j} e^{-\left| \mathbf{r}_j - \mathbf{r}_k \right|^2/2\sigma_j^2} },
    \label{eq:partition}
\end{equation}
where $\mathcal{N}_j$ is the set of ionizable sites within the cutoff of hydrogen $j$, and $\sigma_j$ sets the smoothness of the assignment of hydrogen $j$ to its surrounding ionizable sites.
The formal charge of site $i$ is then
\begin{equation}
    q_i = \sum_{j\in \mathcal{N}_i} n_{j\to i} - n_{i,\mathrm{ref}},
    \label{eq:charge}
\end{equation}
where $\mathcal{N}_i$ is the set of hydrogen atoms within the cutoff of site $i$ and $n_{i,\mathrm{ref}}$ is the number of hydrogen atoms bound to the site in its neutral state, \eg two for a water oxygen.
In this way a hydronium oxygen carries a formal charge close to $+1$, a hydroxide oxygen a charge close to $-1$, and all other oxygen atoms a charge close to zero, with smooth transitions during proton transfer that make the \cv{} differentiable, as required for biasing.
In practice, the \cv{} is implemented as a custom torchscript module made available to \texttt{PLUMED}~\cite{PLUMED,plumed2} through its \texttt{metatomic} interface~\cite{bigi+26jcp}.

Once the formal charges are available, the separation between the products can be defined as a charge-weighted sum of distances from a reference point,
\begin{equation}
    s(\mathbf{R}) = \sum_{i} q_i\, \left|\mathrm{mic}\left(\mathbf{r}_i -\mathbf{r}_{\mathrm{ref}}\right) \right|,
    \label{eq:r-def}
\end{equation}
where the sum runs over the oxygen atoms of the solvent, $\mathrm{mic}(\cdot)$ denotes the minimum-image convention, and $\mathbf{r}_\mathrm{ref}$ is the position of the center of the acid or base, which for the phosphate species is the P atom.
Note that the use of the minimum image is essential: any definition that does not respect the periodicity of the cell would introduce discontinuities in $s$ when an ion crosses a boundary.
To interpret \cref{eq:r-def}, consider the case when a proton is transferred between the solute and a single water molecule, all the solvent oxygen atoms but one carry a negligible charge, and \cref{eq:r-def} reduces to
\begin{math}
    s(\mathbf{R}) = q_{\ce{O^*}}\left|\mathrm{mic}\left(\mathbf{r}_{\ce{O^*}} -\mathbf{r}_{\mathrm{ref}}\right) \right|,
\end{math}
\ie to the distance between the two products, weighted by the sign of the charge of the solvent ion.
This ``signed distance'' is the key to describing the dissociation of amphiprotic species.
An acidic dissociation releases an excess proton into the solvent, so that $q_{\ce{O^*}}>0$ and $s>0$; a basic dissociation creates a hydroxide, so that $q_{\ce{O^*}}<0$ and $s<0$.
The two reactions of an amphiprotic ion such as \ce{H2PO4^-} thus appear on opposite sides of the origin of a one-dimensional free energy profile, 
The undissociated species corresponds to $s\approx 0$, where the sign of $s$ only reflects small fluctuations in the smooth charge assignment, and is common to the two branches.
As the proton diffuses away, the charge is transferred smoothly between solvent molecules and $s$ follows the position of the charge defect, without the need to track a specific atom.

The autoionization of water is a special case, because both products are fluxional and there is no atom that can serve as the reference $\mathbf{r}_\mathrm{ref}$.
Fortunately, the separation between a single pair of opposite charges can be written as a dipole-like expression,
\begin{equation}
    s(\mathbf{R}) = \left|\sum_i q_i\, \mathrm{mic}\left(\mathbf{r}_i-\mathbf{r}_\mathrm{ref}\right)\right|
    \approx \left|\mathrm{mic}\left(\mathbf{r}_+-\mathbf{r}_\mathrm{ref}\right) - \mathrm{mic}\left(\mathbf{r}_--\mathbf{r}_\mathrm{ref}\right)\right|,
    \label{eq:water-cv}
\end{equation}
which reduces to the distance between hydronium ($\mathbf{r}_+$) and hydroxide ($\mathbf{r}_-$) whenever only two sites are charged, and where the reference point only needs to be close enough to both ions for the minimum-image convention to be unambiguous.
We take $\mathbf{r}_\mathrm{ref}$ to be the ``charge center'' of the system, computed for each Cartesian direction $\alpha$ as a circular mean that is compatible with the periodicity of the box,
\begin{equation}
    r_{\mathrm{ref},\alpha}=\frac{L_\alpha}{2\pi}\operatorname{tan}^{-1}\left[\sum_{i}|q_i|\sin\left(\frac{2\pi r_{i,\alpha}}{L_\alpha}\right),\sum_{i}|q_i|\cos\left(\frac{2\pi r_{i,\alpha}}{L_\alpha}\right)\right] \bmod L_\alpha,
    \label{eq:charge-center}
\end{equation}
where $L_\alpha$ is the length of the box along $\alpha$.

One last complication arises for acids and bases whose \pk is comparable with \pkw.
Among the species we consider, the \pka of \ce{HPO4^{2-}} and the \pkb of \ce{H2PO4^-} are only about two \pk units below \pkw, which means that when the bias potential has grown large enough to sample the dissociation of the solute it is also large enough to induce the autoionization of the solvent.
Autoionization events produce a hydronium and a hydroxide that both contribute to \cref{eq:r-def}, and distort the free energy profile of the reaction one is interested in.
To monitor and, if needed, suppress them, we introduce an auxiliary \cv{}
\begin{equation}
    \delta = \sum_{i\in\ce{O}_\mathrm{water}}q_i^2 - \Big(\sum_{i\in\ce{O}_\mathrm{water}}q_i\Big)^2,
    \label{eq:delta-cv}
\end{equation}
which vanishes whenever the solvent contains at most one charged site, as is the case for an acidic or basic dissociation of the solute, and becomes positive when an additional hydronium-hydroxide pair is present.
A one-sided linear restraint with the force constant of \SI{200}{kJ/mol} that activates at $\delta=0.15$ is able to keep the sampling confined to the desired reaction without affecting the associated and dissociated states of the solute.
To further curb the autoionization happening at the associated state, stronger and stricter restraints are needed.
To do so, we first define the charge of the phosphate ion: $Q = \sum_{i\in\ce{O}_\mathrm{phosphate}} \tilde{q}_i$, where $\tilde{q}_i=q_i+5n_{\ce{P}\to i}$, standing for partitioning the $+5$ formal charge carried by the phosphorus atom to the surrounding oxygen atoms.
This allows us to apply stronger linear restraint with the force constant of \SI{500}{kJ/mol} at $\delta=0.05$ to the state where the charge of acid or base has a deviation within $-0.05$ to $0.02$ to its equilibrium state.
These restraints are only applied to the simulations involving \ce{NaH2PO4} or \ce{Na2HPO4}.
For stronger acids and bases, such as \ce{H3PO4} and \ce{Na3PO4}, this restraint is not needed, but we recommend monitoring $\delta$, especially if the bias in metadynamics is left to grow far above the values needed to trigger dissociation.

\subsection{Sampling the dissociation}
\label{sec:sampling}

Given a one-dimensional descriptor, well-tempered metadynamics~\cite{metad,wtm} is the natural choice to accelerate the dissociation, both for the autoionization of water and for the acidic and basic dissociation of a solute.
Gaussian hills are deposited along $s$ and accumulate into a bias potential $V_\mathrm{bias}(s)$ that, in the long-time limit, converges to $-(1-\gamma^{-1})G(s)$, where $\gamma$ is the bias factor and $G(s)$ the free energy as a function of the \cv{}.
In practice, we found that it is hard to converge $G(s)$ quantitatively from the bias alone.
The free energy differences involved are large (up to about $32\,\kT$ for the autoionization of water, according to the experimental value), and in several cases the contact ion pair (\cip) is separated from the solvent-separated pair by kinetic barriers that involve the reorganization of the solvation shell and are not well described by the separation coordinate, so that the bias keeps oscillating around the converged profile.
A second, more fundamental reason to go beyond plain metadynamics is that the quantum corrections we discuss in \cref{sec:nqe} require averages over the associated and dissociated states, and it is much easier to define and compute Boltzmann weights when the bias is a fixed function of the coordinates, although dynamical reweighting of metadynamics is also possible~\cite{tiwa-parr14jpcb,gibe+20jctc}.

For these reasons we use metadynamics only to build a reasonable bias, which is then frozen at a level at which recrossings between the associated and dissociated states have already occurred.
Sampling is then performed with the fixed potential $V(\mathbf{R})+V_\mathrm{bias}(s(\mathbf{R}))$, and the unbiased free energy profile is recovered by reweighting~\cite{torrie1977nonphysical},
\begin{equation}
    G(s) = -\frac{1}{\beta}\ln
    \frac{\left\langle\delta\left(s-s(\mathbf{R})\right)e^{\beta V_\mathrm{bias}\left(s(\mathbf{R})\right)}\right\rangle_\mathrm{bias}}
    {\left\langle e^{\beta V_\mathrm{bias}\left(s(\mathbf{R})\right)}\right\rangle_\mathrm{bias}},
    \label{eq:reweight}
\end{equation}
where $\langle\cdot\rangle_\mathrm{bias}$ indicates an average over the biased ensemble and $\delta(\cdot)$ is the Dirac delta.
Since the frozen bias is only approximately equal to the negative of the free energy, and given the ``hidden" barriers associated with the \cip, we use replica exchange molecular dynamics (\remd) to accelerate the sampling of the fixed-bias ensemble through exchanging eight replicas with different temperatures ranging from \SI{300}{K} to \SI{400}{K}.
Running 10 independent \remd simulations from different initial configurations sampled from the metadynamics simulation provides, in addition, a straightforward estimate of the statistical uncertainty on the resulting averaged \pk, which we report throughout.

The free energy profile $G(s)$ is finally converted into the potential of mean force that enters \cref{eq:equilib-const}.
Since $e^{-\beta G(r)}$ is the probability density of the separation, while $g(r)$ is a density per unit volume, the two differ by the radial Jacobian,
\begin{equation}
    W_\mathrm{sim}(r)=G(r)+k_\mathrm{B}T\ln\left(4\pi r^2\right),
    \label{eq:pmf-from-fes}
\end{equation}
so that in fact $4\pi r^2 e^{-\beta W_\mathrm{sim}(r)}=e^{-\beta G(r)}$ and the integral in \cref{eq:equilib-const} can be evaluated directly on $G(r)$, up to the offset $C$ that is fixed by \cref{eq:offset} on the \pmf.
For the signed dissociation distance of \cref{eq:r-def}, the acidic and basic branches are calibrated separately, each with $r=|s|$ and its own offset $C$, but they share the same associated state. 
The integral in \cref{eq:equilib-const} is therefore extended slightly beyond the origin, so that it includes the whole population of the undissociated species: it runs over $-s_0<s<r_\mathrm{c}$ for the acidic branch and over $-r_\mathrm{c}<s<s_0$ for the basic one, with $s_0=\SI{1}{\angstrom}$.
One should also be careful about the choice of thermodynamic ensemble.
The dissociated state has a different partial molar volume than the associated state, and in a constant-volume simulation of a small box the corresponding change in pressure introduces a small contribution to the free energy difference.
We therefore recommend performing all the sampling in the isothermal-isobaric ensemble, so that the volume can relax and \cref{eq:equilib-const} yields a Gibbs free energy of dissociation, which is what is measured experimentally.
Fixing the volume to the experimental value, which is done commonly to compensate for shortcomings of the DFT reference, is particularly problematic. It leads to an even larger deviation from the volume corresponding to the consistent treatment with the selected \mlip reference, so that one obtain results that might benefit from error cancellation, but fail to capture the true behavior of the underlying electronic-structure method.

\subsection{Computing quantum corrections}
\label{sec:nqe}

Everything discussed so far treats the nuclei as classical particles, which is a poor approximation for a reaction that involves the transfer of a proton.
In principle, one could repeat the enhanced-sampling procedure using \pimd, either depositing the bias on the centroid of the ring polymer or reusing the frozen classical bias, and obtain a quantum free energy profile.
This is, however, very demanding: converging the ring polymer for hydrogen at room temperature requires several tens of replicas, and many of the techniques that reduce this cost, such as the colored-noise thermostats of the PIGLET family~\cite{ceri-mano12prl}, are designed to sample the Boltzmann distribution of the physical system and cannot be used in combination with a bias that alters the target ensemble.
The cost of a quantum \fes has been reported to be more than an order of magnitude larger than that of the corresponding classical calculation~\cite{dasgupta2025nuclear}.

The alternative, which we adopt, exploits the fact that  the quantum correction to the dissociation constant can be computed by thermodynamic integration over the nuclear mass~\cite{ceri-mark13jcp}, where only the centroid kinetic energies of the associated and the dissociated states are required.
Consider a system with Hamiltonian $H(\lambda)=\sum_k \mathbf{p}_k^2/2m_k + V(\mathbf{R})$, in which all nuclear masses are scaled by a common factor, $m_k = \lambda\, m_k^\mathrm{phys}$, and let $A(\lambda)=-\beta^{-1}\ln\operatorname{Tr}\,e^{-\beta H(\lambda)}$ be its free energy.
Since $\partial H/\partial m=-T/m$, where $T$ is the total kinetic energy operator,
\begin{equation}
    \frac{\partial A(\lambda)}{\partial \lambda} = -\frac{\left\langle T\right\rangle_\lambda}{\lambda},
    \label{eq:dadm}
\end{equation}
where $\langle T\rangle_\lambda$ is the quantum expectation value of the kinetic energy for mass $\lambda$.
For $\lambda\to\infty$ the system becomes classical and $\langle T\rangle_\lambda\to d\kT/2$, with $d$ the number of degrees of freedom, so the mass dependence of the classical free energy is entirely contained in the ideal-gas term $-(d/2\beta)\ln \lambda$.
Subtracting this term and integrating from the physical mass to the classical limit, one obtains the quantum correction to the free energy as a convergent integral,
\begin{equation}
    \Delta A_\mathrm{NQE} = A(1)-A_\mathrm{cl}(1) = \int_{1}^{\infty}\frac{\left\langle T\right\rangle_\lambda - d/2\beta}{\lambda}\,\mathrm{d}\lambda.
    \label{eq:ti-mass}
\end{equation}
The integrand decays as $1/\lambda^2$, which makes a direct numerical quadrature inefficient.
With the change of variable $\lambda=1/y^2$, so that $y=1$ corresponds to the physical mass and $y=0$ to the classical limit, \cref{eq:ti-mass} becomes
\begin{equation}
    \Delta A_\mathrm{NQE} = 2\int_{0}^{1}\frac{\left\langle T\right\rangle_{1/y^2} - d/2\beta}{y}\,\mathrm{d}y,
    \label{eq:ti-y}
\end{equation}
whose integrand is smooth, vanishes linearly at $y\to 0$ and can be integrated accurately with a handful of points~\cite{ceri-mark13jcp}.
The same derivation applies in the isothermal-isobaric ensemble, with the Gibbs free energy $G$ in place of $A$, provided that the kinetic energy is averaged at constant pressure.

For the dissociation constant we need the correction to the free energy difference between the dissociated (D) and associated (A) states, and since both states have the same number of degrees of freedom the ideal-gas term cancels, leaving
\begin{equation}
    \Delta\Delta G_\mathrm{NQE} = 2\int_{0}^{1}\frac{\left\langle T\right\rangle^\mathrm{D}_{1/y^2} - \left\langle T\right\rangle^\mathrm{A}_{1/y^2}}{y}\,\mathrm{d}y .
    \label{eq:ddg-nqe}
\end{equation}
The quantum-corrected \pk is then obtained by adding $\Delta\Delta G_\mathrm{NQE}/(\kT\ln 10)$ to the classical value.
Each of the kinetic energy averages in \cref{eq:ddg-nqe} is computed from a \pimd simulation of the corresponding state, using the centroid virial estimator~\cite{herm-bern82jcp}, which converges much faster than the free energy profile itself.
Because no bias along $s$ is needed, these simulations can take full advantage of the acceleration techniques for path integrals, and the associated state is stable on its own; the dissociated state needs to be kept from recombining, which we achieve with a restraint on $s$ that only acts far from the region of interest.
It is worth stressing that the correction obtained in this way is specific to the potential and to the species: it depends on the change in the zero-point energy and in the delocalization of the proton between the two states, and there is no reason to expect a universal value that could be transferred from one system to another, or from one level of theory to another.

\section{Computing acid/base equilibria with ML potentials}
\label{sec:mlip}

The workflow discussed in \cref{sec:theory} requires an amount of sampling that is out of reach for direct \textit{ab initio} \md: long reactive trajectories for each species, several independent replicas, and path integral simulations for a grid of masses.
Machine-learning potentials bring this within reach at a fixed, and comparatively small, cost, and remove any incentive to cut corners in the protocol.
Neural-network potentials trained on \dft data have already been used to compute the autoionization of water with the DeePMD or Behler-Parrinello frameworks~\cite{deepmd,behl-parr07prl, calegari2023probing,liu2023mechanistic}, and, in combination with explicit path integral simulations, to quantify the nuclear quantum correction to \pkw~\cite{dasgupta2025nuclear}.
The agreement with experiment that has been reported for purely classical calculations~\cite{calegari2023probing,liu2023mechanistic} should however be considered in the light of the fact that the classical limit of \pkw, as extrapolated from the experimental isotope dependence, is about three units above the experimental value~\cite{ceri+16cr}: the errors of the potential and the neglect of \nqe{s} can compensate each other, and a transferable and predictive protocol must include the quantum correction explicitly.

The choice of the reference level of theory deserves a comment.
Foundation models are usually trained on datasets computed at the generalized-gradient (\gga) or meta-\gga level, which are affordable for millions of structures but are not accurate enough for a quantitative treatment of hydrogen bonding and proton transfer in water.
We use the hybrid functional revPBE0~\cite{PBE1,PBE2,PBE3} with the D3 dispersion correction~\cite{dftd3}, which has been shown repeatedly to describe the structure, dynamics and thermodynamics of liquid water very well when combined with a quantum treatment of the nuclei~\cite{marsalekQuantumDynamicsSpectroscopy2017,chengInitioThermodynamicsLiquid2019,kapil2022first}, as well as aqueous solutions of ions~\cite{gomez2024neural,NaCl-revPBE0,falkner2026dfttrainedneuralnetworkpotentials,garcia2026protonation}.
We should stress that a recent, careful study indicates that part of this success may itself be due to a cancellation between the errors of the functional and those of the basis set~\cite{beck2026more}; the protocol we describe is agnostic to the level of theory.

\subsection{MLIP architecture and validation}
\label{sec:model}

We use the Point Edge Transformer (\pet) architecture~\cite{pet}, a graph-neural-network potential that is not constrained to be rotationally equivariant by construction, but is made so by a smooth, exact symmetrization scheme, and which has been found to give excellent accuracy across diverse chemical systems.
Our base model is PET-MAD~\cite{pet-mad}, in the PET-MAD-1.5 version trained on the MAD-1.5 dataset of highly diverse structures~\cite{PET-MAD-1.5-2026}.
The model has three graph neural network (\gnn) layers, and we set a fixed cutoff of \SI{8.0}{\angstrom}, resulting in a receptive field of \SI{24.0}{\angstrom}.
Fine-tuning a foundation model to a different level of theory requires only a small fraction of the data that would be needed to train from scratch~\cite{mattersim,tompa2026finetuning,radova2025fine}, and the base model can be used to sample the relevant chemical space in the first round of an active-learning loop, which avoids expensive \textit{ab initio} \md at the outset.

The model used in this work was fine-tuned on a dataset labeled at the level of revPBE0-D3 used in Ref.~\citenum{chengInitioThermodynamicsLiquid2019}.
The fine-tuning was carried out in an active-learning-like fashion.
The initial dataset combined the water dataset from Ref.~\citenum{chengInitioThermodynamicsLiquid2019} with a manually curated set of aqueous \ce{NaCl}, \ce{NaOH} and \ce{HCl} configurations, which introduce the sodium counterions of the phosphate salts together with the hydroxide and hydronium products of proton transfer.
Metadynamics simulations of \ce{H3PO4}, \ce{NaH2PO4}, \ce{Na2HPO4} and \ce{Na3PO4}, each solvated in a box of 128 water molecules, were used to sample the rare proton-transfer events between the solute and water and between water molecules, and \pimd simulations were added to cover the more delocalized configurations that classical sampling does not reach.
After a few rounds of training and simulation the dataset contained 6099 structures, and the final model reached a mean absolute error (\mae) of \SI{1.07}{meV/atom} on the energies and of \SI{19.2}{meV/\angstrom} on the forces.
All simulations were performed with the \texttt{metatomic} interface~\cite{bigi+26jcp} to \texttt{i-PI}~\cite{i-pi}, and \texttt{PLUMED}~\cite{PLUMED,plumed2}; simulation details are given in the Supporting Information.

\subsection{Water autoionization}
\label{sec:water}

\begin{figure}
    \centering
    \includegraphics[width=0.7\linewidth]{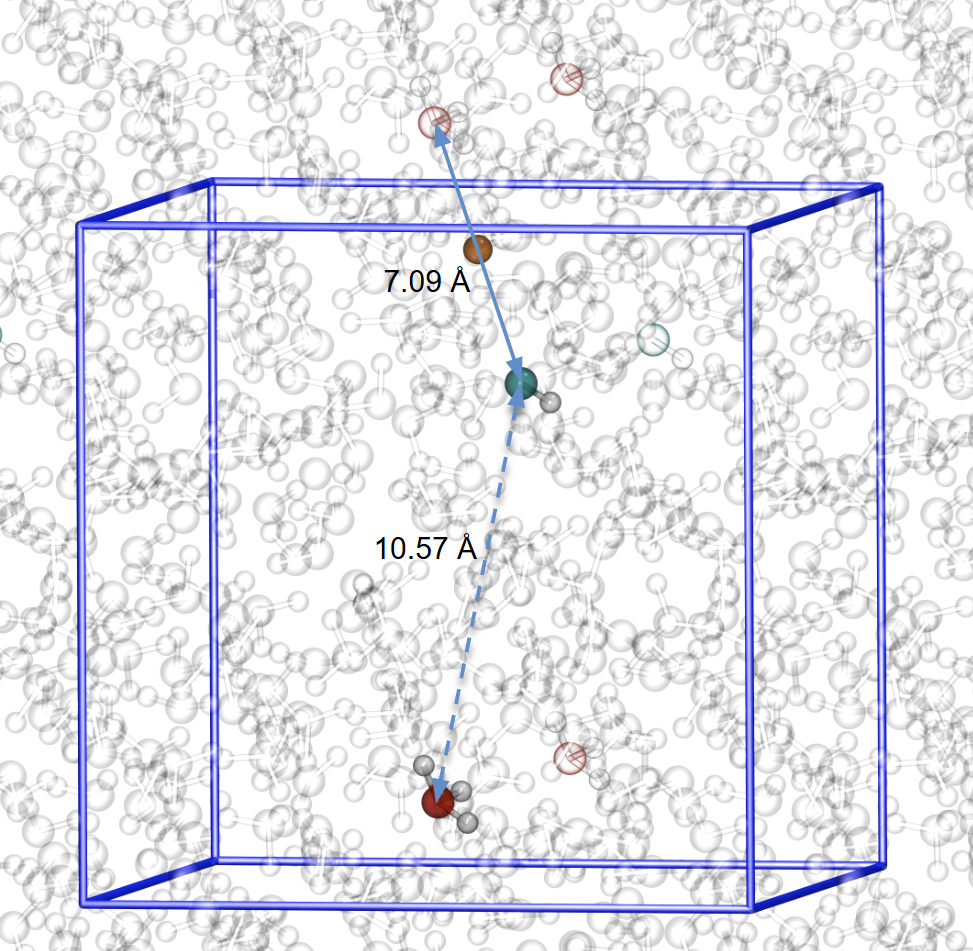}
    \caption{Snapshot of a dissociated configuration from the biased simulation of water, with the oxygen atoms colored according to their formal charge $q_i$ (\cref{eq:charge}). The hydronium is colored in red, and hydroxide is colored in cyan. Their image atoms are colored in the same way but in transparent. The charge center $\mathbf{r}_\mathrm{ref}$ is colored in orange. The remaining background water molecules are shown transparently. The blue frame represents the simulation box. The solid arrow stands for the correct ion-ion distance captured by the \cv, and the dashed arrow stands for the distance calculated without the treatment of \cref{eq:charge-center}.}
    \label{fig:water-snapshot}
\end{figure}

We discuss the autoionization of water in some detail, because it is the simplest case and because it makes it possible to compare with a large body of previous work.
The system contains 128 water molecules, and the dissociation distance is the dipole-like separation of \cref{eq:water-cv}.
All simulations are conducted at \SI{300}{K} and \SI{1}{bar}.
\Cref{fig:water-snapshot} shows a snapshot of a dissociated configuration, with the formal charges of \cref{eq:charge} highlighted, which illustrates how the charge of the two defects is localized on one, or shared between a few, oxygen atoms during the Grotthuss diffusion, while all other water molecules remain essentially neutral.

\begin{figure}
    \centering
    \includegraphics[width=1.0\linewidth]{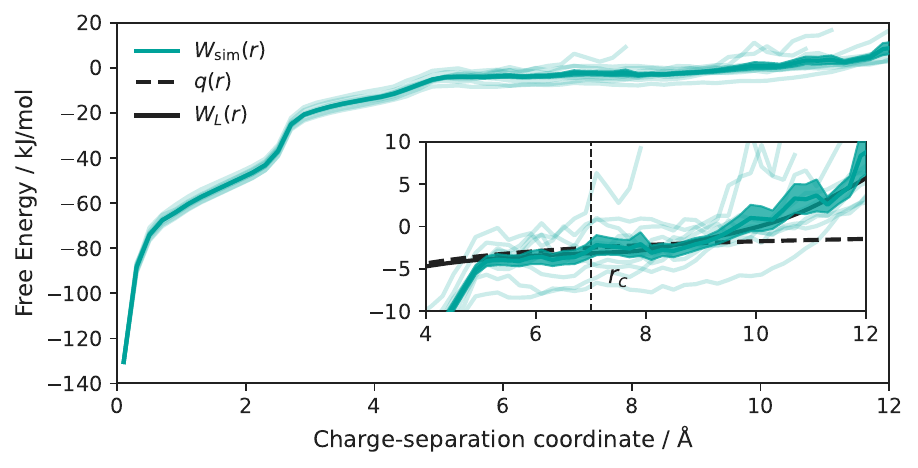}
    \caption{Classical free energy profile $W_\textrm{sim}$ of liquid water as a function of the charge-separation coordinate of \cref{eq:water-cv}, from reweighted fixed-bias replica-exchange simulations. Thin lines correspond to independent runs started from different configurations, and the thick line to their average. Inset: dissociated region, compared with the screened Coulomb interaction of two point charges in the thermodynamic limit, $q(r)$ (dashed), and in a periodic box of the same size including the accessible-shell correction, $W_L(r)$ (solid black); the vertical line marks $r_\mathrm{c}$.}
    \label{fig:water-fes}
\end{figure}

\Cref{fig:water-fes} shows the classical free energy profile along $s$, obtained from the reweighted fixed-bias \remd simulations.
The profile rises steeply for $s<\SI{0.6}{\angstrom}$, then climbs more gently up to about \SI{2.3}{\angstrom}, which corresponds to the transfer of a proton to a neighboring molecule and to a hydronium-hydroxide contact ion pair.
Between \SI{2.3}{\angstrom} and \SI{3.0}{\angstrom} the free energy jumps by about \SI{20}{kJ/mol}, reaching \SI{110}{kJ/mol} when the ions are separated by one water molecule, and after \SI{7.0}{\angstrom} it reaches a matching region where the \pmf is only weakly modulated by the residual Coulomb attraction.
We take this region as the onset of the dissociated state, and set $r_\mathrm{c}=\SI{7.0}{\angstrom}$ in \cref{eq:equilib-const}.
The spread among the ten independent \remd runs shown as thin lines in \cref{fig:water-fes} gives a direct estimate of the statistical uncertainty, which is of the order of a few kJ/mol in the dissociated region and translates into an uncertainty of $\pm 0.37$ \pk units on the classical \pkw.
The density of the system also changes during the dissociation.
From the associated state to the dissociated state, the density increases by more than $1\%$ (\cref{SI-fig:water-density}), underscoring the need for NpT sampling.

The calibration of the offset $C$ is illustrated in \cref{fig:water-fes}, which compares the simulated profile in the dissociated region with the screened Coulomb interaction of two point charges, in the thermodynamic limit and in a periodic box of the same size as the simulation cell.
The two reference curves differ visibly over the range of separations that are accessible in a box of 128 molecules, which shows that finite-size effects are not negligible at this system size, and that calibrating against the periodic reference is necessary to obtain a result that can be compared with experiment.

\begin{figure}
    \centering
    \includegraphics[width=0.80\linewidth]{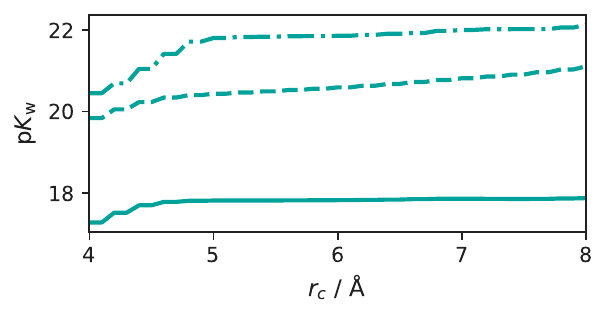}
    \caption{Comparison of different \pk calculation methods and their sensitivities to $r_\mathrm{c}$. Solid lines, dashed lines, and dash-dotted lines represent \pk calculated through the reversible work theorem, integral-over-states, and free energy difference between minima of two states. The integral-over-states method calculates \pk through $-\log_{10}(\int_{r_\mathrm{c}}^{10}\exp\left[-\beta W_\mathrm{sim}(r)\right]/\int_0^{r_\mathrm{c}}\exp\left[-\beta W_\mathrm{sim}(r)\right]$, and free energy minima method calculates through $-\log_{10}(\exp-\beta\left[\min_{r_\mathrm{c}<r<10}W_\mathrm{sim}(r) - W_\mathrm{sim}(0)\right])$.}
    \label{fig:rc-sensitivity}
\end{figure}

With the profile calibrated, \cref{eq:equilib-const} yields a classical $\pkw=17.86\pm0.37$, which corresponds to a Gibbs free energy of dissociation of \SI{102.6}{kJ/mol}.
This value compares well with the estimate of about 17 for the classical limit of \pkw that was obtained by extrapolating the experimental isotope dependence~\cite{ceri+16cr}.
It is instructive to compare with what other estimators would give on the very same profile.
Taking the difference between the free energy minimum of the associated state and the plateau minimum in \cref{fig:water-fes} gives \SI{126.5}{kJ/mol}, which is \SI{23.9}{kJ/mol}, or 4.2 \pk units, higher than the reversible-work result.
The discrepancy is due to the fact that the free energy at a point of the profile carries the Jacobian of the \cv{} and the entropy of the volume of configurations that map onto it, and does not correspond to the standard state that defines \pkw.
\cref{fig:rc-sensitivity} compares different \pk calculation methods, where several \pk units discrepancy is shown between different methods.
Differences of this size are larger than the spread of the values reported in the literature for water, and larger than any plausible error of the potential, so it is essential to be explicit about the estimator when comparing calculations with each other and with experiment.

\begin{figure}
    \centering
    \includegraphics[width=0.7\linewidth]{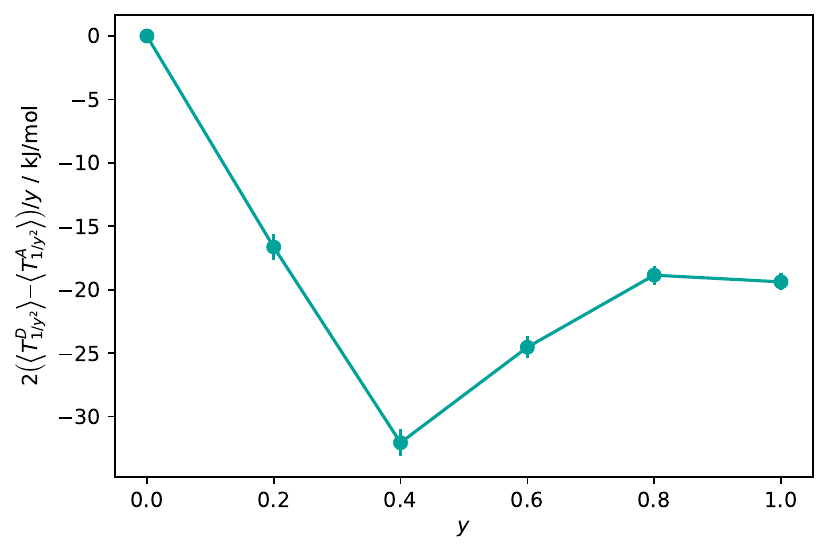}
    \caption{Integrand of the mass thermodynamic integration, \cref{eq:ddg-nqe}, for the autoionization of water: difference between the centroid-virial kinetic energies of the dissociated and associated states, divided by $y$. $y=0$ corresponds to the classical limit, where the integrand vanishes identically, and $y=1$ to the physical masses. Error bars indicate the statistical uncertainty on each point. The area under the curve is $\Delta\Delta G_\mathrm{NQE}$.}
    \label{fig:water-ti}
\end{figure}

\Cref{fig:water-ti} shows the integrand of \cref{eq:ddg-nqe}, \ie the difference in centroid-virial kinetic energy between the dissociated and the associated state, as a function of the mass-scaling variable $y$.
The kinetic energy of the dissociated state is lower at all masses, consistent with the conversion of one covalent O-H bond to a delocalized proton.
The integrand is relatively smooth and can be integrated with five \pimd simulations, giving $\Delta\Delta G_\mathrm{NQE} = -20.36 \pm 0.38\ \mathrm{kJ/mol}$, \ie a correction of $-3.54$ \pk units that brings the value to $\pkw=14.31\pm0.37$, in good agreement with the experimental value of 13.93.
The correction is very large, and there is no way to obtain a quantitative \pkw from a simulation that treats the nuclei classically, other than by a compensation of errors.

Our quantum-corrected \pkw is consistent with the value of $13.71\pm0.16$ reported by Dasgupta \textit{et al.}~\cite{dasgupta2025nuclear} from explicit \pimd simulations with a different potential, at a fraction of the cost: the total cost to obtain quantum-corrected results based on existing classical result is about 120 million single-point force evaluations in our case, against 1.76 billion in Ref.~\citenum{dasgupta2025nuclear}.
The agreement on the final number conceals, however, substantial differences in the ingredients.
Their classical \pkw is $19.44\pm0.11$, and the quantum correction is 5.73 \pk units, against 17.86 and 3.54 in our case.
Since the estimators differ (integration over states versus reversible work) and so do the potentials, it is not possible to attribute the difference to a single cause, but the comparison makes it clear that the quantum correction is not a universal constant that can be added to any classical result.
In the same vein, classical calculations that reported values close to experiment, such as 14.7~\cite{calegari2023probing} and 14.14~\cite{liu2023mechanistic}, should be read as indicating a compensation between the errors of the potential energy surface and the neglect of \nqe{s}, rather than as evidence that a classical treatment is sufficient.
This is also the reason why the agreement we obtain for water cannot, by itself, be taken as a validation of the workflow, and why we test it on a family of related dissociations in the next section.

\subsection{Phosphoric acid}
\label{sec:phosphate}

Phosphoric acid provides an ideal test, because its three acidic dissociations span ten \pk units, and because the intermediate species \ce{H2PO4^-} and \ce{HPO4^{2-}} are amphiprotic, so that a total of six dissociation constants, three \pka and three \pkb, can be computed and checked for consistency with \cref{eq:conjugate-pair}.
We simulate \ce{H3PO4}, \ce{NaH2PO4}, \ce{Na2HPO4} and \ce{Na3PO4}, each in a box of 128 water molecules, with sodium counterions to keep the cell neutral, and we use the signed dissociation distance of \cref{eq:r-def} with the P atom as the reference point, together with the restraint on $\delta$ to suppress the autoionization of the solvent where needed (\cref{sec:cv}).
No other adjustment of the overall protocol was needed to move from water to the phosphate species.
It should be noted that one solute in 128 water molecules corresponds to a concentration of about \SI{0.4}{mol/L}, while the experimental values are at infinite dilution.
Furthermore, the radial distribution functions of \ce{P}-\ce{Na} shown in \cref{SI-fig:phosphate-rdf} indicate that the phosphate ions interact with the counterions.
The effects from these phosphate ion-counterion interactions are not taken into account.
Experiments conducted at high background salt concentrations show decreased \pka values for the three dissociations of phosphoric acid~\cite{peter1989dissociation} up to 1.5 \pk units under the same ionic strengths comparable to those of the corresponding simulations.
Even though our simulations contain no salt other than the phosphate species and their counterions, these observations suggest that the computed \pk values may deviate from their infinite-dilution limit.

\begin{figure}
    \centering
    \includegraphics[width=0.8\linewidth]{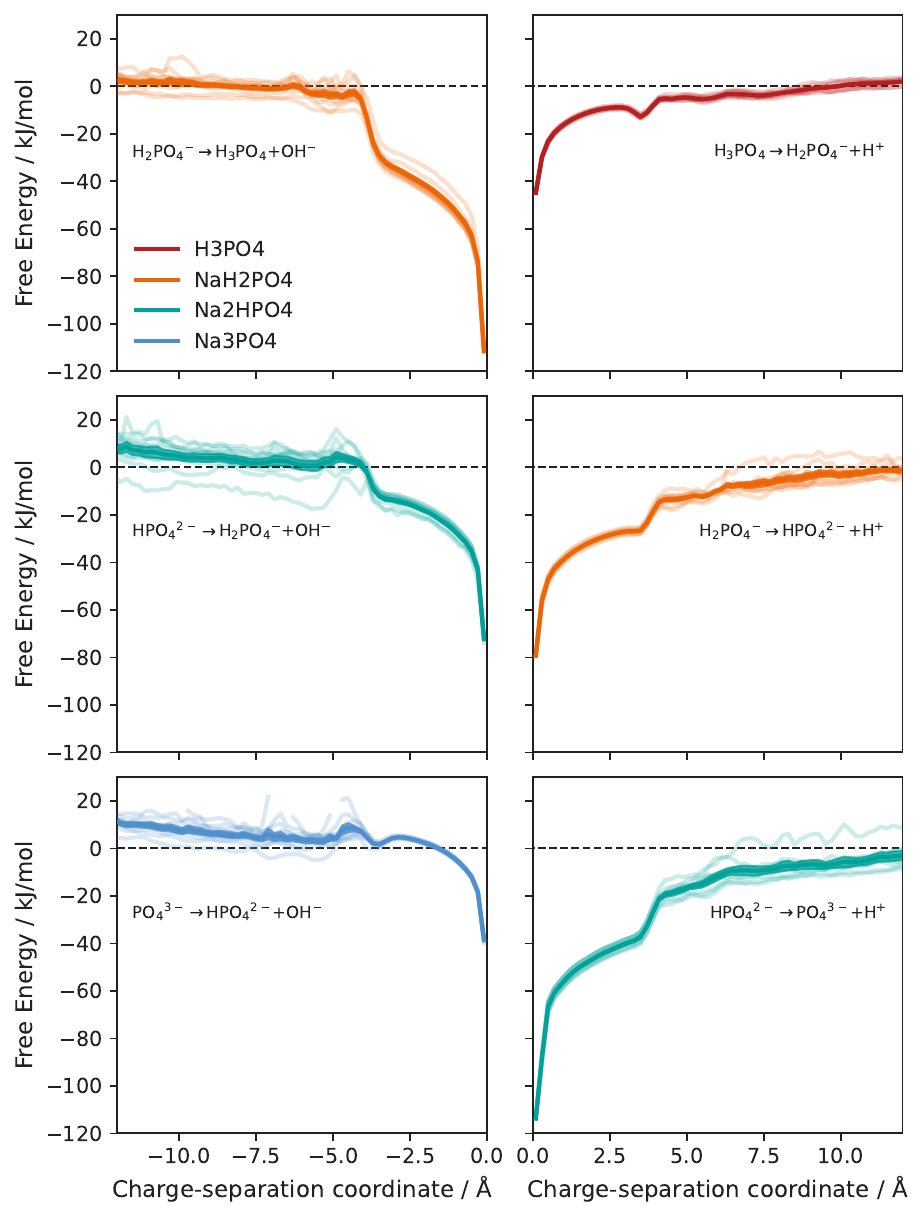}
    \caption{Classical free energy profiles $W_\textrm{sim}$ of the four phosphate species as a function of the signed charge-separation coordinate of \cref{eq:r-def}. Positive values correspond to the acidic dissociation, negative values to the basic dissociation. Thin lines are the reweighted profiles of individual fixed-bias replica-exchange runs, ten for each species, started from different initial configurations; thick lines are their average, and the shaded bands indicate one standard deviation.}
    \label{fig:fess}
\end{figure}

\Cref{fig:fess} shows the free energy profiles of the four species along the signed coordinate.
Each row corresponds to a conjugate acid-base pair: the left panels show the basic dissociations, and the right panels the acidic dissociation of the corresponding conjugate acid.
The ordering of the free energy differences between the dissociated and associated regions reproduces the expected trend: each successive dissociation is more difficult.
The profiles also reveal the changing nature of the contact ion pair along the ladder.
For \ce{H3PO4} and \ce{Na3PO4}, a metastable \cip is visible as a shoulder between the associated state and the dissociated plateau, separated from the latter by a small barrier.
The stability of the \cip decreases gradually with each successive dissociation stage: the barrier between the \cip and the associated state becomes vanishingly small for the acidic dissociation of \ce{NaH2PO4} and the basic dissociation of \ce{Na2HPO4}, and disappears altogether for the acidic dissociation of \ce{Na2HPO4} and the basic dissociation of \ce{NaH2PO4}, while the barrier towards the dissociated state grows.
It is precisely for these cases that the fixed-bias replica-exchange sampling makes a difference, suggesting that the separation coordinate alone does not describe the reorganization of the solvation shell that accompanies the escape from the \cip, and that the barrier is present even though it is not visible on the dissociation profile.

\begin{table}
    \caption{Classical and quantum-corrected dissociation constants of water and of the phosphate species, compared with experiment. $\Delta_\mathrm{NQE}$ is the correction from \cref{eq:ddg-nqe} in \pk units. Experimental values for the phosphate species are from Ref.~\citenum{goldberg2002thermodynamic} and converted to the values at \SI{300}{K} with the provided free energy differences and enthalpy differences; the experimental \pkb values are obtained from the \pka of the conjugate acid and the experimental \pkw at \SI{300}{K}~\cite{IAPWS_R11-24}. }
    \label{tab:pka}
    \begin{tabular}{llrrrr}
    \hline
    Species & Reaction & Classical & $\Delta_\mathrm{NQE}$ & Quantum & Expt. \\
    \hline
    \ce{H2O}      & \pkw & $17.86\pm0.37$ & $-3.54\pm0.07$ & $14.31\pm0.37$ & 13.93 \\
    \ce{H3PO4}    & \pka & $3.00\pm0.06$  & $-0.90\pm0.06$ & $2.10\pm0.09$  & 2.16 \\
    \ce{NaH2PO4}  & \pka & $9.00\pm0.22$  & $-2.12\pm0.06$ & $6.87\pm0.23$  & 7.19 \\
    \ce{Na2HPO4}  & \pka & $15.06\pm0.28$ & $-3.05\pm0.06$ & $12.02\pm0.29$ & 12.33 \\
    \ce{NaH2PO4}  & \pkb & $15.22\pm0.13$ & $-2.38\pm0.06$ & $12.85\pm0.14$ & 11.77 \\
    \ce{Na2HPO4}  & \pkb & $8.60\pm0.15$  & $-1.42\pm0.06$ & $7.17\pm0.16$  & 6.74 \\
    \ce{Na3PO4}   & \pkb & $2.29\pm0.20$  & $-0.11\pm0.06$ & $2.18\pm0.21$  & 1.60 \\
    \hline
    \end{tabular}
\end{table}

\begin{figure}
    \centering
    \includegraphics[width=0.6\linewidth]{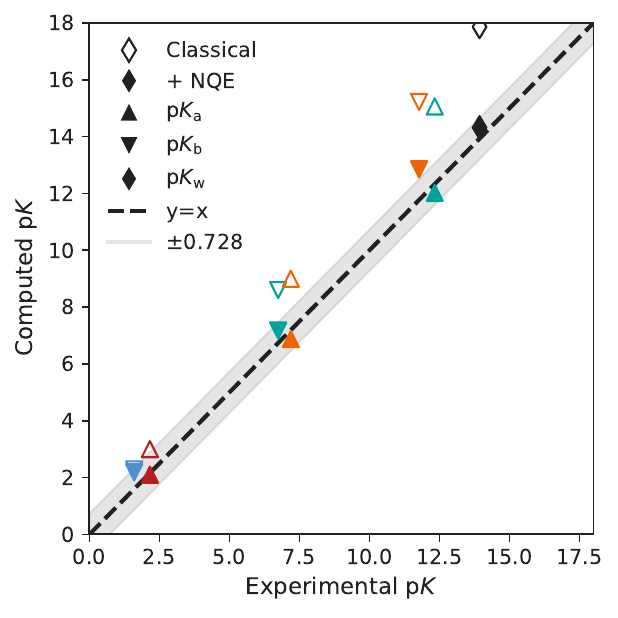}
    \caption{Computed \pka (upward triangles) and \pkb (downward triangles) of the phosphate species, and \pkw (diamond) against the experimental values~\cite{goldberg2002thermodynamic,IAPWS_R11-24}. Empty symbols are the classical values from \cref{eq:equilib-const}, and full symbols include the quantum correction of \cref{eq:ddg-nqe}. Colors identify the species as in \cref{fig:fess}. The shaded band around the diagonal corresponds to a deviation of $\pm\SI{1}{kcal/mol}$ in the free energy of dissociation.}
    \label{fig:pka}
\end{figure}

The classical and quantum-corrected dissociation constants are collected in \cref{tab:pka} and compared with experiment in \cref{fig:pka}.
The classical values are systematically too large, by 2.19 \pk units on average, consistent with the overestimation of \pkw.
The quantum correction lowers all the \pk values, but by very different amounts: from 0.11 units for the \pkb of \ce{Na3PO4} to 3.54 units for the \pkw, with a rough trend for the correction to grow with the \pk itself.
After the correction, the mean absolute error with respect to the experiment drops from 2.19 to 0.45 \pk units, \ie well within the \SI{1}{kcal/mol} that is conventionally taken as chemical accuracy and that corresponds to about 0.7 \pk units at room temperature.
The variability of the correction across the ladder of the same acid, together with the comparison with the literature discussed in \cref{sec:water}, confirms that \nqe{s} must be computed explicitly for each dissociation, and that they cannot be accounted for by a constant shift, not even within the same potential.

The six constants of phosphate dissociations are not independent, and \cref{eq:conjugate-pair} provides a stringent test of internal consistency that does not rely on experiment.
Summing the \pka of each acid and the \pkb of its conjugate base gives three independent estimates of \pkw: 14.95 for the \ce{H3PO4}/\ce{H2PO4^-} pair, 14.05 for \ce{H2PO4^-}/\ce{HPO4^{2-}}, and 14.20 for \ce{HPO4^{2-}}/\ce{PO4^{3-}}.
The latter two agree with the directly computed value of 14.31 to within the combined uncertainties, while the first deviates by 0.6 \pk units.
Repeating the exercise with the classical values gives 18.22, 17.60 and 17.35 against the direct 17.86.
These classical values show comparable deviations without a clear outlier, so that the discrepancy for the first pair results from the combination of a deviation of about 0.4 units in the classical profiles and a difference of about 0.3 units in the quantum corrections.
The outlier among the quantum sums is the \ce{H3PO4}/\ce{H2PO4^-} pair, which involves the \pkb of \ce{NaH2PO4}, \ie the value that deviates most from experiment in \cref{fig:pka}.
It is worth noting that the basic dissociation of \ce{H2PO4^-} is the only reaction in the set whose products, neutral \ce{H3PO4} and \ce{OH^-}, do not interact through a charge-charge term, so that the calibration of \cref{eq:offset} cannot be applied to it, and the lack of a proper calibration may explain the large deviation.
Whatever the origin, this kind of consistency check is available for any polyprotic acid and for any conjugate pair, and we recommend it as a routine diagnostic, since agreement with experiment can always be the result of a compensation of errors, whereas a violation of \cref{eq:conjugate-pair} cannot.

\section{Conclusions}
\label{sec:conclusions}

We have presented a step-by-step protocol to compute absolute acid and base dissociation constants from atomistic simulations, and we have applied it to the autoionization of water and to the entire dissociation ladder of phosphoric acid.
The protocol relies on the reversible work theorem to define the dissociation constant in terms of a potential of mean force whose asymptotic normalization is fixed by matching to the Coulomb interaction of point charges in the same periodic cell, on a single signed distance based on smoothly assigned formal charges, on metadynamics followed by fixed-bias replica-exchange sampling to obtain converged Boltzmann weights, and on mass thermodynamic integration to evaluate the nuclear quantum correction separately for each dissociation.
None of the steps is specific to phosphate: moving from water to the six dissociations of the phosphate species required no change in the definitions of the order parameters or in the analysis; only the metadynamics parameters were adapted to the magnitude of the dissociation free energy.

The main practical limitation is the choice and optimization of the potential.
Foundation models make it possible to obtain a reactive \mlip for a new solute with a few thousand reference calculations, but the level of theory to which the model is fine-tuned determines the final accuracy, and it is unlikely that a \gga functional would give quantitative results.
We used revPBE0-D3, which has a good track record for water, but even this choice may benefit from a cancellation of errors~\cite{beck2026more}, and the deviation we observe for the \pkb of \ce{H2PO4^-} may be a manifestation of the residual error of the reference method, of the model, or of the presence of counterions in a small box, none of which can be disentangled without further reference calculations.
The consistency of conjugate pairs through \cref{eq:conjugate-pair} offers at least an internal check that does not depend on experiment, and that we recommend to include in every calculation of this kind.

As better quantum references and better \mlip{s} become available, this kind of rigorous modeling workflow will become generally applicable, and we expect it would be relatively easy to extend it to other modeling scenarios, such as the simulation of polyelectrolytes, the determination of the protonation states of protein residues, or the acid-base chemistry of surface sites.

\begin{acknowledgement}
We thank Matthias Kellner for sharing the \ce{NaCl} dataset and Alessandro Forina for helpful discussions.
QX acknowledges support from the COST Action CA22154 - DAEMON. MC and PL acknowledge support from the European Research Council (ERC) under the European Union's Horizon 2020 research and innovation programme (Grant Agreement No. 101001890-FIAMMA).
SL was supported by the Shanghai Jiaotong University ``Jing'e" Undergraduate Overseas Training Program.
CM and MC acknowledge support from the NCCR Separations, a National Centre of Competence in Research funded by the Swiss National Science Foundation (grant number 229280).
PL acknowledges funding from the German Research Foundation (DFG) under Germany’s Excellence Strategy - EXC 2089/2 – 390776260. 
Computing resources were provided by Scientific IT \& Application Support (SCITAS) of EPFL, the Swiss National Supercomputing Centre under project no. lp133.
\end{acknowledgement}

\begin{suppinfo}
Simulation details for the fine-tuning of the potential, the enhanced-sampling and replica-exchange simulations, the path-integral simulations and the thermodynamic integration, as well as the evaluation of the Coulomb calibration curves.
A tutorial example of the calculation of \pkw will be uploaded to https://github.com/lab-cosmo/atomistic-cookbook upon publication, and data and models will be made available on the Materials Cloud archive.
\end{suppinfo}

\providecommand{\latin}[1]{#1}
\makeatletter
\providecommand{\doi}
  {\begingroup\let\do\@makeother\dospecials
  \catcode`\{=1 \catcode`\}=2 \doi@aux}
\providecommand{\doi@aux}[1]{\endgroup\texttt{#1}}
\makeatother
\providecommand*\mcitethebibliography{\thebibliography}
\csname @ifundefined\endcsname{endmcitethebibliography}
  {\let\endmcitethebibliography\endthebibliography}{}

\clearpage


\section*{Supplementary material}

# Supporting Information: Supporting Information for: Quantitative modeling of acid-base equilibria

Qianjun Xu, Simin Liu, Cesare Malosso, Philip Loche, and Michele Ceriotti*

*Laboratory of Computational Science and Modeling, Institut des Matériaux, École Polytechnique Fédérale de Lausanne, 1015 Lausanne, Switzerland*

E-mail: michele.ceriotti@epfl.ch

## S1 Calculation of dissociations and autoionization constants

The single molecule partition function of species $i$ is

$$\begin{aligned} q_i &= \frac{1}{h^{3n}} \int_{V^n} d\mathbf{r} \int d\mathbf{p} e^{-\beta H(\mathbf{r},\mathbf{p})} \\ &= \frac{1}{h^{3n}} \int_{V^n} \prod_k d\mathbf{r}_k \int \prod_k d\mathbf{p}_k e^{-\beta\left[\sum_k \frac{\mathbf{p}_k^2}{2m_k} + U(\mathbf{r}_1,\mathbf{r}_2,\ldots,\mathbf{r}_n)\right]}, \end{aligned} \tag{S1}$$

where $V$ is the volume of the system, $\mathbf{r}_k$ and $\mathbf{p}_k$ are the position and momentum of the $k$-th atom, and $n$ is the number of atoms in the molecule.

For the equilibrium

$$\mathrm{HA} \rightleftharpoons \mathrm{H}^+ + \mathrm{A}^- \tag{S2}$$

that we are interested in, we assume that $H^+$ and $A^-$ both act as a single particle, i.e. their

partition function is

$$q_i = \frac{1}{h^3}\int_V d\mathbf{r}\int d\mathbf{p} e^{-\beta\left[\frac{\mathbf{p}^2}{2m_i}+U(\mathbf{r})\right]}$$
$$= V/\Lambda_i^3, \tag{S3}$$

where $m_i$ is the mass of species $i$, and $\Lambda_i = h/\sqrt{2m_i\pi/\beta}$ is the thermal wavelength of species $i$; and particle HA is a bound state of the particle $H^+$ and $A^-$, whose partition function is

$$q_i = \frac{1}{h^6}\int_{V^2} d\mathbf{r}_1 d\mathbf{r}_2 \int d\mathbf{p}_1 d\mathbf{p}_2 e^{-\beta\left[\frac{\mathbf{p}_1^2}{2m_1}+\frac{\mathbf{p}_2^2}{2m_2}+U(\mathbf{r}_1,\mathbf{r}_2)\right]}$$
$$= \frac{1}{h^6}\int_V d\mathbf{r}_{\mathrm{com}}\int_{V_{\mathrm{rel}}} d\mathbf{r}_{\mathrm{rel}}\int d\mathbf{p}_{\mathrm{com}}\int d\mathbf{p}_{\mathrm{rel}} e^{-\beta\left(\frac{\mathbf{p}_{\mathrm{com}}^2}{2m_{\mathrm{tot}}}+\frac{\mathbf{p}_{\mathrm{rel}}^2}{2m_{\mathrm{red}}}+U(\mathbf{r}_{\mathrm{rel}})\right)}$$
$$= \frac{V}{\Lambda_{\mathrm{HA,tot}}^3} q_i^{\mathrm{int}}, \tag{S4}$$

where the transformation

$$\begin{cases} \mathbf{r}_{\mathrm{com}} & = (m_1\mathbf{r}_1 + m_2\mathbf{r}_2)/(m_1+m_2) \\ \mathbf{r}_{\mathrm{rel}} & = \mathbf{r}_1 - \mathbf{r}_2 \\ \mathbf{p}_{\mathrm{com}} & = \mathbf{p}_1 + \mathbf{p}_2 \\ \mathbf{p}_{\mathrm{rel}} & = (m_2\mathbf{p}_1 - m_1\mathbf{p}_2)/(m_1+m_2) \end{cases} \tag{S5}$$

is applied, $m_{\mathrm{tot}} = m_1 + m_2$ is the total mass, $m_{\mathrm{red}} = \frac{m_1 m_2}{m_1+m_2}$ is the reduced mass, $\Lambda_{\mathrm{HA,tot}} = h/\sqrt{2m_{\mathrm{tot}}\pi/\beta}$ is the thermal wavelength of the bound state, and

$$q_i^{\mathrm{int}} = \frac{1}{h^3}\int_{V_{\mathrm{rel}}} d\mathbf{r}_{\mathrm{rel}}\int d\mathbf{p}_{\mathrm{rel}} \exp\left[-\beta\left(\frac{\mathbf{p}_{\mathrm{rel}}^2}{2m_{\mathrm{red}}}+U(\mathbf{r}_{\mathrm{rel}})\right)\right] \tag{S6}$$

is the intramolecular partition function. Here, we assume that the potential only depends on the relative position of these two particles.

For a system containing $N_{\mathrm{H}}$ particle $H^+$, $N_{\mathrm{A}}$ particle $A^-$, and $N_{\mathrm{HA}}$ particle HA, the

partition function of the system is

$$Q = \frac{1}{N_{\mathrm{H}}!N_{\mathrm{A}}!N_{\mathrm{HA}}!} q_{\mathrm{H}}^{N_{\mathrm{H}}} q_{\mathrm{A}}^{N_{\mathrm{A}}} q_{\mathrm{HA}}^{N_{\mathrm{HA}}}. \tag{S7}$$

The Helmholtz free energy is

$$\begin{aligned} A &= -k_{\mathrm{B}}T \ln Q \\ &= -k_{\mathrm{B}}T \sum_i N_i \ln q_i - \ln N_i! \\ &\approx -k_{\mathrm{B}}T \sum_i N_i \ln q_i + N_i - N_i \ln N_i. \end{aligned} \tag{S8}$$

The chemical potential of species $i$ is

$$\begin{aligned} \mu_i &= (\frac{\partial A}{\partial N_i})_{V,T,N_{j\neq i}} \\ &= -k_{\mathrm{B}}T(\ln \frac{q_i}{N_i}). \end{aligned} \tag{S9}$$

For clarity, we write the partition function as the product of the translational partition function $q_i^{\mathrm{trans}} = V/\Lambda_i^3$ and the intramolecular partition function $q_i^{\mathrm{int}}$, that is, $q_i = q_i^{\mathrm{trans}} q_i^{\mathrm{int}}$:

$$\begin{aligned} \mu_i &= -k_{\mathrm{B}}T(\ln q_i^{\mathrm{int}} + \ln q_i^{\mathrm{trans}} - \ln N_i) \\ &= -k_{\mathrm{B}}T(\ln q_i^{\mathrm{int}} + \ln V - 3\ln \Lambda_i - \ln n_i V) \\ &= -k_{\mathrm{B}}T(\ln q_i^{\mathrm{int}} - 3\ln \Lambda_i - \ln c_i/c^\circ - \ln N_{Av}c^\circ) \\ &= \mu_i^\circ + k_{\mathrm{B}}T \ln c_i/c^\circ, \end{aligned} \tag{S10}$$

where $n_i$ is the number density of species $i$, $c_i$ is the molar concentration of species $i$, $N_{Av}$ is the Avogadro constant, $\mu_i^\circ$ is the standard chemical potential of species $i$, which is

$$\mu_i^\circ = -k_{\mathrm{B}}T(\ln q_i^{\mathrm{int}}/\Lambda_i^3 - \ln N_{Av}c^\circ), \tag{S11}$$

and $c^\circ = 1\,\mathrm{mol/L}$ is the standard molar concentration.

At equilibrium of the reaction $\mathrm{HA} \rightleftharpoons \mathrm{H}^+ + \mathrm{A}^-$,

$$\begin{aligned} \mu_{\mathrm{HA}} &= \mu_{\mathrm{H}} + \mu_{\mathrm{A}} \\ \mu^\circ_{\mathrm{HA}} - \mu^\circ_{\mathrm{H}} - \mu^\circ_{\mathrm{A}} &= k_{\mathrm{B}}T \ln \frac{c_{\mathrm{H}} c_{\mathrm{A}}}{c_{\mathrm{HA}} c^\circ} \\ k_{\mathrm{B}}T \ln \frac{(q^{\mathrm{int}}_{\mathrm{H}}/\Lambda^3_{\mathrm{H}})(q^{\mathrm{int}}_{\mathrm{A}}/\Lambda^3_{\mathrm{A}})}{(q^{\mathrm{int}}_{\mathrm{HA}}/\Lambda^3_{\mathrm{HA,tot}})} \frac{1}{N_{Av} c^\circ} &= k_{\mathrm{B}}T \ln K^\circ_d, \end{aligned} \tag{S12}$$

where $K^\circ_d$ is the standard dissociation constant:

$$K^\circ_d = \frac{1}{n^\circ} \frac{(q^{\mathrm{int}}_{\mathrm{H}}/\Lambda^3_{\mathrm{H}})(q^{\mathrm{int}}_{\mathrm{A}}/\Lambda^3_{\mathrm{A}})}{(q^{\mathrm{int}}_{\mathrm{HA}}/\Lambda^3_{\mathrm{HA,tot}})}, \tag{S13}$$

and $n^\circ = N_{Av} c^\circ = 1/1660\ \text{Å}^{-3}$ is the standard number density. Substituting the intramolecular partition functions into Equation (S13), we have

$$\begin{aligned} K^\circ_d &= \frac{1}{n^\circ} \frac{\Lambda^3_{\mathrm{HA,tot}}}{\Lambda^3_{\mathrm{H}} \Lambda^3_{\mathrm{A}}} \left[ \frac{1}{h^3} \int_{V_{\mathrm{rel}}} d\mathbf{r}_{\mathrm{rel}} \int d\mathbf{p}_{\mathrm{rel}} e^{\left[ -\beta \left( \frac{\mathbf{p}^2_{\mathrm{rel}}}{2m_{\mathrm{red}}} + U(\mathbf{r}_{\mathrm{rel}}) \right) \right]} \right]^{-1} \\ &= \frac{1}{n^\circ} \frac{\Lambda^3_{\mathrm{HA,tot}}}{\Lambda^3_{\mathrm{H}} \Lambda^3_{\mathrm{A}}} \frac{h^3}{(2m_{\mathrm{red}}\pi/\beta)^{3/2}} \left[ \int_{V_{\mathrm{rel}}} d\mathbf{r}_{\mathrm{rel}} e^{-\beta U(\mathbf{r}_{\mathrm{rel}})} \right]^{-1} \\ &= \frac{1}{n^\circ} \left[ \int_{V_{\mathrm{rel}}} d\mathbf{r}_{\mathrm{rel}} e^{-\beta U(\mathbf{r}_{\mathrm{rel}})} \right]^{-1} \end{aligned} \tag{S14}$$

where $\frac{h}{(2m_{\mathrm{red}}\pi/\beta)^{1/2}}$ can be considered as the thermal wavelength of the reduced mass $m_{\mathrm{red}}$. Here, it is hard to handle $V_{\mathrm{rel}}$. We consider a cutoff $r_{\mathrm{c}}$, for $|\mathbf{r}| > r_{\mathrm{c}}$, the compound HA is no longer considered as bound, but two free particles $\mathrm{H}^+$ and $\mathrm{A}^-$, and $\{\mathbf{r} \,|\, |\mathbf{r}| < r_{\mathrm{c}}\} \subset V_{\mathrm{rel}}$ holds. As a result,

$$K^\circ_d = \frac{1}{n^\circ} \left[ 4\pi \int_0^{r_{\mathrm{c}}} dr_{\mathrm{HA}} r^2_{\mathrm{HA}} e^{-\beta U(r_{\mathrm{HA}})} \right]^{-1}. \tag{S15}$$

This is the result for acid/base dissociation. For the water autoionization, the definition of the standard chemical potential $\mu^\circ_{\mathrm{w}}$ is different from those above for solutes. We start

from the chemical potential expression Equation (S10):

$$
\begin{aligned}
\mu_{\mathrm{w}} &= -k_{\mathrm{B}}T(\ln q_{\mathrm{w}}^{\mathrm{int}}/\Lambda_{\mathrm{w}}^3 - \ln N_{Av}c^\circ) + k_{\mathrm{B}}T \ln c/c^\circ \\
&= -k_{\mathrm{B}}T(\ln q_{\mathrm{w}}^{\mathrm{int}}/\Lambda_{\mathrm{w}}^3 - \ln N_{Av}c^\circ) + k_{\mathrm{B}}T \ln x_{\mathrm{w}}\rho/M/c^\circ \\
&= -k_{\mathrm{B}}T(\ln q_{\mathrm{w}}^{\mathrm{int}}/\Lambda_{\mathrm{w}}^3 + \ln V_{\mathrm{H_2O}}) + k_{\mathrm{B}}T \ln x_{\mathrm{w}} \\
&= \mu_{\mathrm{w}}^\circ + k_{\mathrm{B}}T \ln x_{\mathrm{w}}, \qquad \text{(S16)}
\end{aligned}
$$

where $x_{\mathrm{w}}$ represents the mole fraction of water in the system, $\rho$ and $M$ represents the density and mean molar mass of the system, and $V_{\mathrm{H_2O}}$ stands for the volume occupied by one water molecule. We use the equilibrium condition again:

$$
\begin{aligned}
\mu_{\mathrm{w}} =& \mu_{\mathrm{H^+}} + \mu_{\mathrm{OH^-}} \\
-k_{\mathrm{B}}T(\ln q_{\mathrm{w}}^{\mathrm{int}}/\Lambda_{\mathrm{w}}^3 + \ln V_{\mathrm{H_2O}}) =& -k_{\mathrm{B}}T(\ln q_{\mathrm{H^+}}^{\mathrm{int}}/\Lambda_{\mathrm{H^+}}^3 - \ln N_{Av}c^\circ) + k_{\mathrm{B}}T \ln(c_{\mathrm{H^+}}/c^\circ) \\
& - k_{\mathrm{B}}T(\ln q_{\mathrm{OH^-}}^{\mathrm{int}}/\Lambda_{\mathrm{OH^-}}^3 - \ln N_{Av}c^\circ) + k_{\mathrm{B}}T \ln(c_{\mathrm{OH^-}}/c^\circ) \\
\ln\left(\frac{q_{\mathrm{H^+}}^{\mathrm{int}}/\Lambda_{\mathrm{H^+}}^3 q_{\mathrm{OH^-}}^{\mathrm{int}}/\Lambda_{\mathrm{OH^-}}^3}{q_{\mathrm{w}}^{\mathrm{int}}/\Lambda_{\mathrm{w}}^3} \cdot \frac{1}{n^{\circ 2} V_{\mathrm{H_2O}}}\right) =& \ln \frac{c_{\mathrm{H^+}} c_{\mathrm{OH^-}}}{c^{\circ 2}}. \qquad \text{(S17)}
\end{aligned}
$$

Thus, the equilibrium constant of water autoionization is:

$$
\begin{aligned}
K_{\mathrm{w}} &= \frac{1}{n^{\circ 2} V_{\mathrm{H_2O}}} \frac{q_{\mathrm{H^+}}^{\mathrm{int}}/\Lambda_{\mathrm{H^+}}^3 q_{\mathrm{OH^-}}^{\mathrm{int}}/\Lambda_{\mathrm{OH^-}}^3}{q_{\mathrm{w}}^{\mathrm{int}}/\Lambda_{\mathrm{w}}^3} \\
&= \frac{1}{n^{\circ 2} V_{\mathrm{H_2O}}} \left[4\pi \int_0^{r_{\mathrm{c}}} dr_{\mathrm{H_2O}} r_{\mathrm{H_2O}}^2 e^{-\beta U(r_{\mathrm{H_2O}})}\right]^{-1}, \qquad \text{(S18)}
\end{aligned}
$$

where $r_{\mathrm{H_2O}}$ represents the distance between the hydronium and hydroxide. Our method, however, allows every water molecule in the system to dissociate. Therefore, the potential of mean force computed from the simulation cannot be directly interpreted as $U(r_{\mathrm{H_2O}})$, and further treatment is required.

We start with the ratio of the probabilities of the system being in the dissociated state, $P_1$, and in the associated state, $P_0 = \int_0^{r_{\mathrm{c}}} p(r)dr$, where $p(r)$ is the probability density of the

system along $r$. This ratio equals the ratio of the partition functions of the dissociated and associated states:

$$
\begin{aligned}
P_1/P_0 = Q_1/Q_0 &= \frac{q_{\mathrm{H_2O}}^{(N-1)} q_{\mathrm{H^+}} q_{\mathrm{OH^-}}/(N-1)!}{q_{\mathrm{H_2O}}^{N}/N!} \\
&= \frac{q_{\mathrm{H^+}} q_{\mathrm{OH^-}} N}{q_{\mathrm{H_2O}}} \\
&= NV \frac{q_{\mathrm{H^+}}^{\mathrm{int}}/\Lambda_{\mathrm{H^+}}^3 q_{\mathrm{OH^-}}^{\mathrm{int}}/\Lambda_{\mathrm{OH^-}}^3}{q_{\mathrm{H_2O}}^{\mathrm{int}}/\Lambda_{\mathrm{H_2O}}^3} \\
&= (n^\circ V)^2 K_\mathrm{w}.
\end{aligned}
\tag{S19}
$$

Equation (S18) is used in the last step.

The probability density function $p(r)$ is related to the radial distribution function (RDF):

$$p(r) = g(r) p_{\mathrm{ideal}}(r), \tag{S20}$$

where $g(r)$ is the RDF. $p_{\mathrm{ideal}}(r)$ is the probability density function of two particles distributed as ideal gas. The equation $p_{\mathrm{ideal}}(r)dr = P_1 \cdot 4\pi r^2 dr/V$ holds, meaning the probability of the system being in the dissociated state and the dissociation species are separated by $r$. With the reversible work theorem $g(r) = \exp[-\beta W(r)]$,

$$p(r)dr = P_1 \cdot 4\pi r^2 \exp[-\beta W(r)] dr/V, \tag{S21}$$

where $W(r)$ is the potential of mean force computed from the simulation.

As a result,

$$P_0 = \int_0^{r_\mathrm{c}} p(r)dr = \int_0^{r_\mathrm{c}} P_1 \cdot 4\pi r^2 \exp[-\beta W(r)] dr/V, \tag{S22}$$

which is

$$P_1/P_0 = V \cdot \left\{ \int_0^{r_\mathrm{c}} 4\pi r^2 \exp\left[-\beta W(r)\right] dr \right\}^{-1} = (n^\circ V)^2 K_\mathrm{w}. \tag{S23}$$

Therefore,

$$K_{\mathrm{w}} = \frac{1}{n^{\circ 2} V} \left\{ \int_0^{r_{\mathrm{c}}} 4\pi r^2 \exp\left[-\beta W(r)\right] dr \right\}^{-1}. \tag{S24}$$

# S2 Fine-tuning of the potential

**Reference calculations.** All reference energies and forces were computed with the revPBE0 hybrid functional[S1–S3] and the D3 dispersion correction,[S4] using the TZV2P basis set with the CP2K software.[S5] Computational details of the calculations are the same as those used in Marsalek and Markland[S6], Cheng et al.[S7].

**Model fine-tuning.** The initial dataset combined the water dataset with a manually curated set of aqueous NaCl, NaOH and HCl configurations. Metadynamics simulations of $H_3PO_4$, $NaH_2PO_4$, $Na_2HPO_4$ and $Na_3PO_4$, each solvated in 128 water molecules, and PIMD simulations were used to harvest configurations that were then labeled at the reference level of theory and added to the training set, in an iterative fashion. The final dataset contained 6099 structures and was split randomly to get the training set and the test set, containing 90% and 10% structures, respectively. The fine-tuned model reached a MAE of 1.07 meV/atom on energies and 19.2 meV/Å on forces on the test set.

**Model validation.** To check that the model reproduces the density of the reference method, we computed the density of water from NpT simulations. The result is $0.937 \pm 0.001$ g/cm$^3$, consistent with the result of classical nuclei of revPBE0-D3 reported in Ref. S7. To further validate the fine-tuned model, we calculated the RDFs of water and phosphates. These RDFs are calculated from 1 ns long unbiased NpT MD simulations conducted at 300 K and 1 bar.

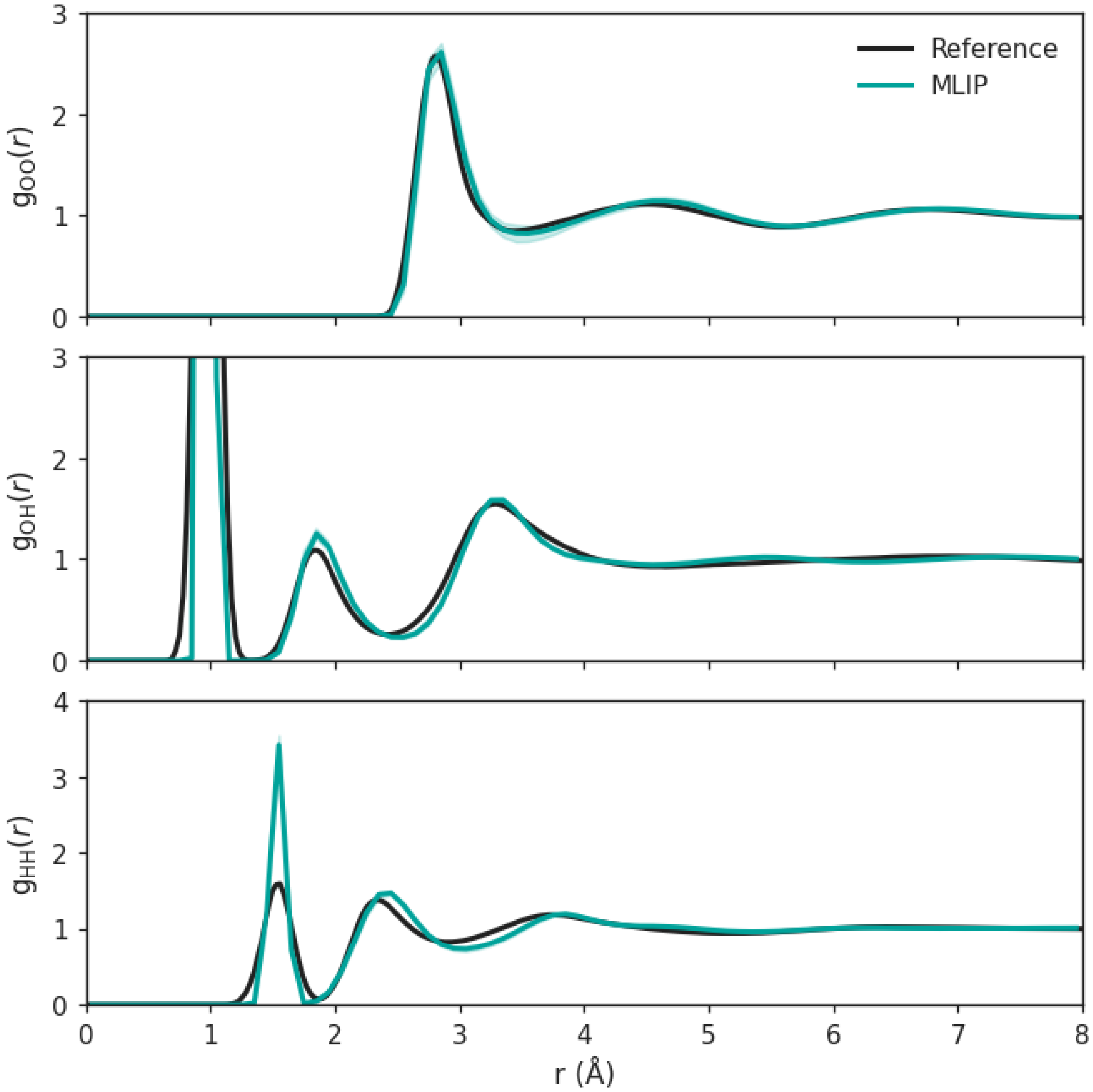


Figure S1: The O-O, O-H, and H-H RDFs of water. Experimental values are collected from Ref. S8 (O-O) and Ref. S9 (O-H and H-H). The shaded areas represent the model uncertainty, which is computed through ensemble-member-reweighting[S10,S11] enabled by shallow ensemble.[S12]

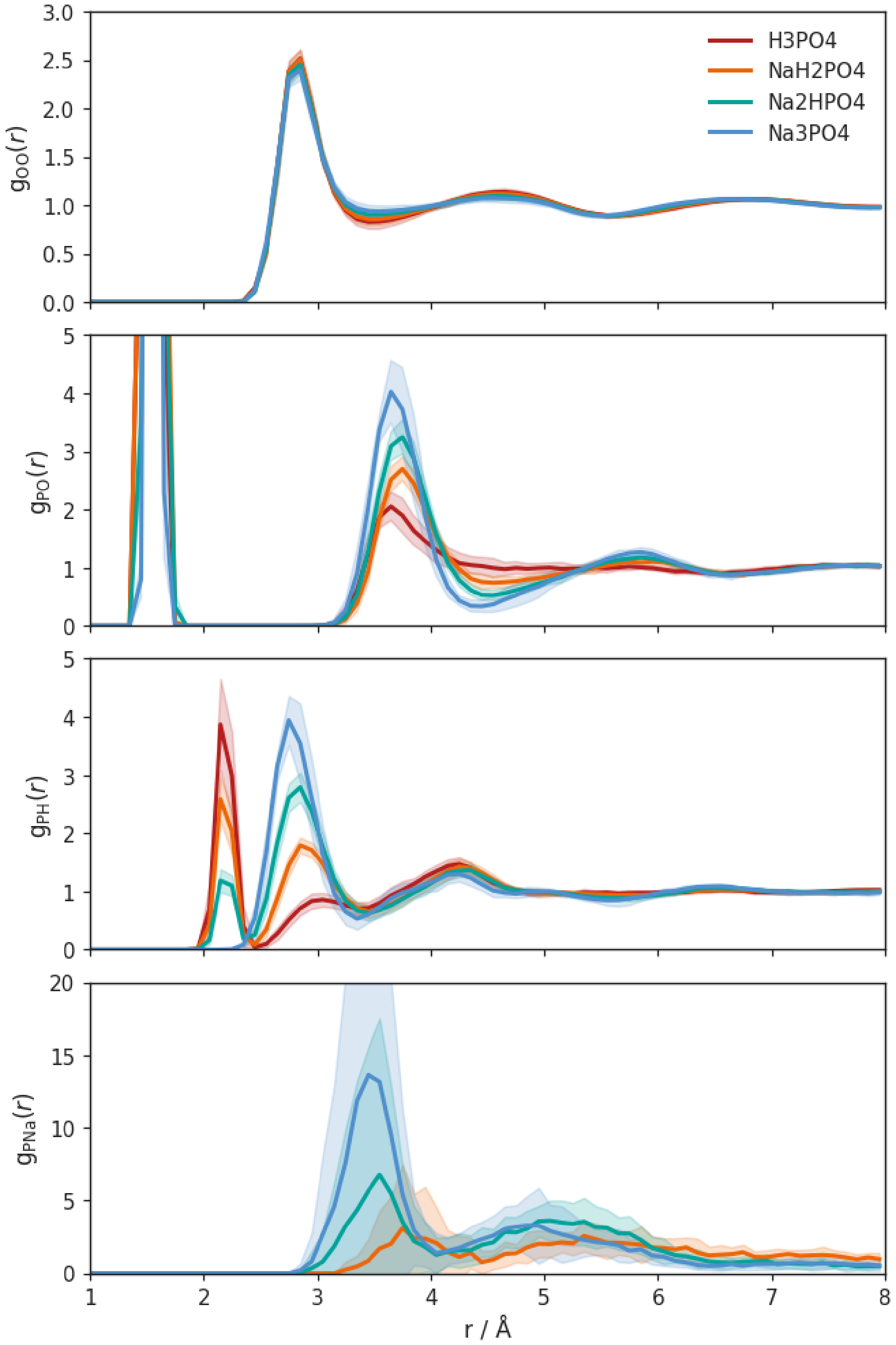


Figure S2: The O-O, P-O, P-H, and P-Na RDFs of the four phosphate species studied in this work. The shaded areas are computed through the same way introduced in Figure S1.

## S3 Dissociation distance

The cutoff for the neighbor list calculation in the calculation of CV is 3.5 Å. The smoothness of the assignment is set as 0.3 Å for all hydrogen atoms in the system and 1.0 Å for the phosphorus atom. The ionizable sites are defined to be all the O atoms in the simulation systems, while only the O atoms in the solvent are taken into account in the calculation of the separation distance.

## S4 Enhanced sampling and replica exchange

The simulations described here are all conducted with `i-PI` v3.3.0, `PLUMED` v2.10, and `metatomic-torch` v0.1.16. All the trajectories, if not otherwise specified, are integrated with a time step of 0.5 fs at 300 K with the stochastic velocity rescaling thermostat[S13] with a time constant of 20 fs and 1 bar with the Bussi-Zykova-Parrinello barostat[S14] with a time constant of 200 fs.

**Metadynamics.** Well-tempered metadynamics[S15,S16] was performed along the signed distance $s$ to build the bias potential, which was frozen at the end of the run. The simulation of water started from a cubic box of an edge length of 15.6482 Å containing 128 water molecules. The simulations of phosphate species started from a rhombic dodecahedron box of an edge length of 17.659 Å containing 128 water molecules and either $H_3PO_4$, $Na^+$ and ${H_2PO_4}^-$, 2 $Na^+$ and ${HPO_4}^{2-}$, or 3 $Na^+$ and ${PO_4}^{3-}$. The parameters used in metadynamics vary for different systems, since the difficulty of dissociating different species is different. The parameters are listed in Table S1. For the water system, one-sided harmonic potentials are placed at $-2$ Å (only as a safeguard, since $s \geq 0$ by construction) and 12 Å. For the phosphate systems, one-sided harmonic potentials are placed at $-14$ Å and 14 Å. These walls prevent the simulations from exploring unnecessarily large values of $|s|$, such as double dissociation.

Table S1: Parameters of the metadynamics runs. The $\sigma$ of the Gaussian bias potential is 0.3 Å. The deposition stride of the bias potentials is 0.5 ps. The remaining species-dependent parameters are listed below.

| Species | Gaussian height / $k_\mathrm{B}T$ | Bias factor | Simulation length / ns |
|---|---|---|---|
| $H_2O$ | 1.0 | 12 | 2.7 |
| $H_3PO_4$ | 0.5 | 5 | 2.2 |
| $NaH_2PO_4$ | 0.8 | 8 | 4.1 |
| $Na_2HPO_4$ | 0.8 | 8 | 4.1 |
| $Na_3PO_4$ | 0.5 | 10 | 4.1 |

**Fixed-bias replica exchange.** Sampling with the frozen bias was performed with replica-exchange molecular dynamics, using ten independent runs started from different configurations for each species; the free energy profiles were obtained by reweighting.[S17] During the simulation, the restricting one-sided potentials mentioned above are applied. Temperatures are exchanged between eight replicas, spanning from 300 K to 400 K: 300 K, 313 K, 326 K, 339 K, 354 K, 368 K, 384 K, and 400 K. Exchanges are attempted every 50 steps. The total trajectory length is 4.3 ns for water, 2.8 ns for $H_3PO_4$, 2.6 ns for $NaH_2PO_4$, 2.4 ns for $Na_2HPO_4$, 2.3 ns for $Na_3PO_4$.

**Density change along the dissociation path.** We monitored the density along the dissociation path. The density of water increased by at most 1.6% after dissociation, as shown in Figure S3.

# S5 Path integral simulations and thermodynamic integration

The quantum correction was computed by thermodynamic integration over the mass (Section "Computing quantum corrections" of the main text), with the centroid-virial kinetic energy estimator.[S18,S19] Five PIMD simulations at $y$ = 0.2, 0.4, 0.6, 0.8 and 1.0, with all nuclear masses scaled by the common factor $1/y^2$, were performed for each of the associated and

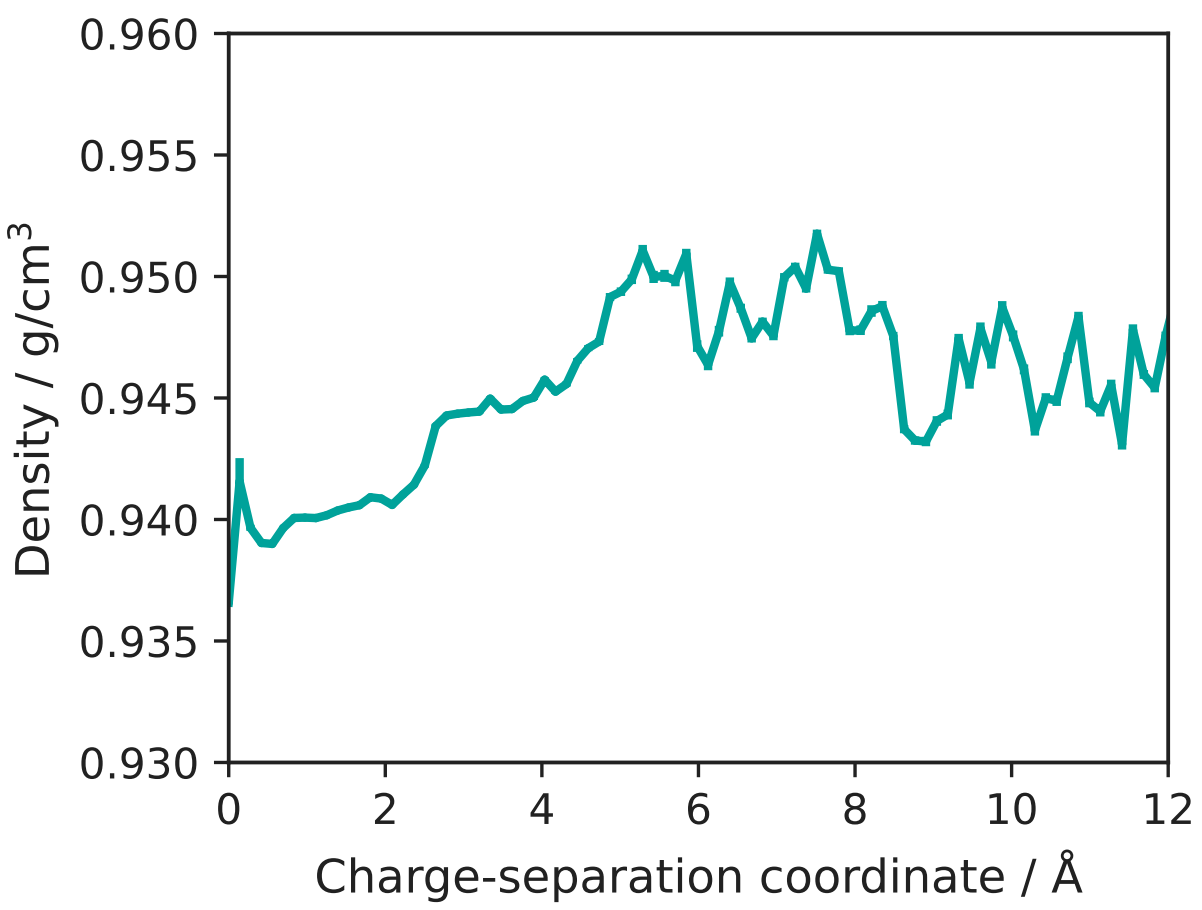


Figure S3: Conditional average of the density of water along the dissociation path, obtained by binning the frames according to their charge-separation coordinate and averaging the density within each bin. These frames are collected from the trajectories sampled by REMD simulations at 300 K.

dissociated states of all the seven dissociations studied here, and the integral was evaluated by quadrature including the point $y = 0$, where the integrand vanishes. The systems are confined in a certain state through one-sided harmonic potentials with a force constant of $5\,\text{kJ/mol/Å}^2$. The associated states are confined by potentials at 3 Å and −3 Å. The acidic and basic dissociated states are confined by potentials at 5 Å and −5 Å, respectively. The PIMD simulations use 32 beads, with all other settings as in Section S4, except for the thermostat, which is a white-noise Langevin thermostat acting on the normal modes of the ring polymer, with a local Langevin thermostat with a time constant of 20 fs acting on the centroid.[S20] Each simulation is at least 170 ps long.

# S6 Calculation of the dissociation constants

**Neutralizing-background correction.** The Wigner self-energy $E_W$ is calculated through `torchpme` package.[S21] The system is set up with a periodic box of the same size and shape as the simulation cell. The dielectric constant is set as 80, the room temperature value. A single point charge of +1 is placed at the origin of the box, and the electrostatic potential

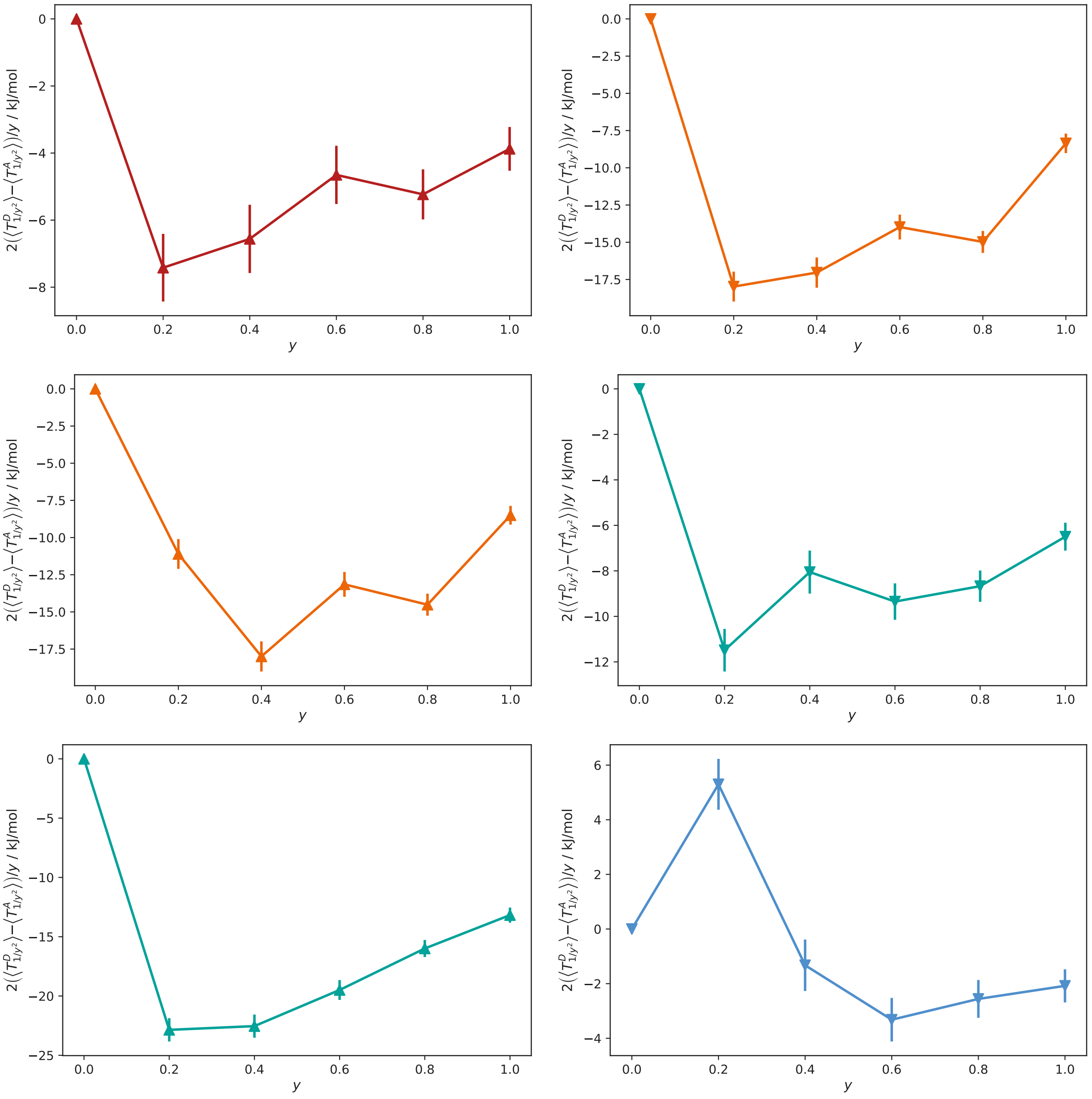


Figure S4: Integrand of the mass thermodynamic integration for the phosphate dissociations, similar to Figure 4. The left column shows the integrands of the acidic dissociations of $H_3PO_4$, $NaH_2PO_4$, and $Na_2HPO_4$, and the right column shows the integrands of the basic dissociations of $NaH_2PO_4$, $Na_2HPO_4$, and $Na_3PO_4$, respectively.

of this periodic system is computed by Ewald summation. The resulting potential energy of the system is the $E_{\mathrm{W}}$.

**Evaluation of the dissociation constants.** The dissociation constants were evaluated with the reversible work theorem (Section 2.1 of the main text), with the cutoff $r_{\mathrm{c}}$ set at 7.0 Å for every species. For the phosphate species, the integral over the associated state runs from $s = -1$ Å to 7 Å for the acidic dissociations and from $-7$ Å to 1 Å for the basic ones, so that the population of the undissociated species, which fluctuates around $s = 0$, is fully included in both cases. The offset $C$ was obtained by matching the simulated potential of mean force to the electrostatic interaction of two point charges in a periodic box of the same size, computed with the `torchpme` package.[S21] It should be noted that the closed form of the Jacobian-related term explained in the Section "What are we computing" is not directly applicable to the dodecahedral boxes. In practice, this term is taken into account when conducting the `torchpme` calculations, through placing one point charge at the origin point and the other at the grid points across the box. The dielectric constant of water is chosen as the room temperature value, 80.

**Sensitivity test of the p$K$ to the choice of $r_{\mathrm{c}}$** We tested the sensitivity of the p$K$ values to the choice of the cutoff radius $r_{\mathrm{c}}$. The upper limit of the range used to compute the offset $C$ is fixed at 10 Å, 4 Å below the restricting potential. The cutoff radius is varied from 4 Å to 8 Å. As shown in Figure S5, the reversible-work p$K$ values are stable, with no system varying by more than one p$K$ unit.

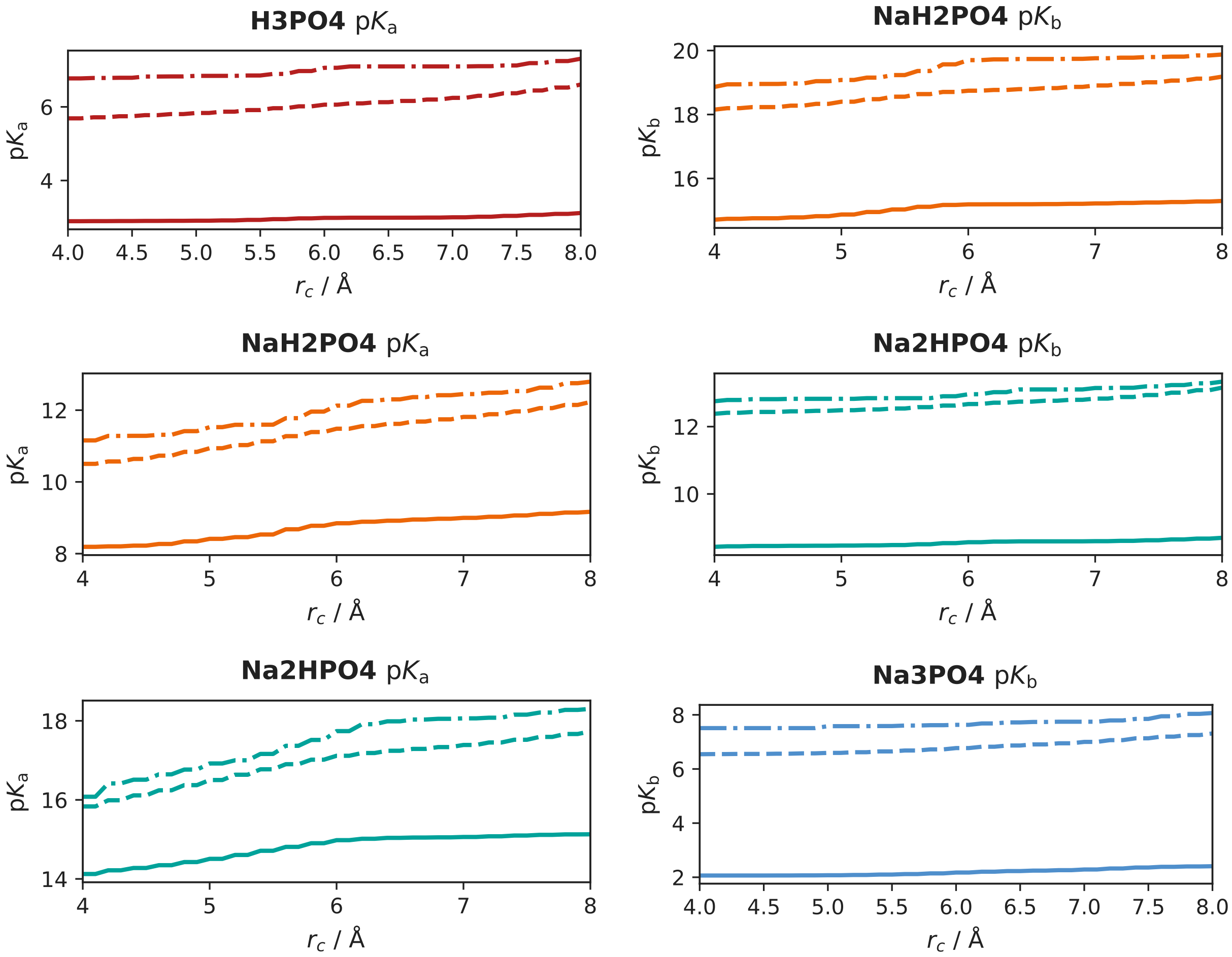


Figure S5: Sensitivity of the phosphate p$K_a$ and p$K_b$ values to the choice of $r_c$. Solid, dashed, and dash-dotted lines correspond to p$K$ values computed with the reversible work theorem, by integrating over the associated and dissociated states, and from the free energy difference between the minima of the two states, respectively. The p$K$ values computed with the reversible work theorem vary by at most 0.12 p$K$ units between 6 Å and 8 Å.



\end{document}